%% file: main.tex
\documentclass[twocolumn, twocolappendix, usenatbib, useAMS]{aastex631}
\usepackage{times, amsmath, amssymb, amsfonts, latexsym}
\usepackage{fleqn, multirow, graphicx, color, enumitem}
\usepackage{array, mathtools, subfigure, esint,CJK}
\usepackage{soul, hyperref}
\usepackage{tikz}
\usetikzlibrary{arrows,shapes,chains}
\newcommand{\uat}[2]{\href{http://astrothesaurus.org/uat/#2}{#1 (#2)}}

\def\mpch{\,{h^{-1} {\rm Mpc}}}          \def\hmpc{\,{h {\rm Mpc}^{-1}}}
\def\dd{{\rm d}} 
\makeatletter
\@for\next:={int,iint,iiint,iiiint,dotsint,oint,oiint,sqint,sqiint,
	ointctrclockwise,ointclockwise,varointclockwise,varointctrclockwise,
	fint,varoiint,landupint,landdownint}\do{%
	\expandafter\edef\csname\next\endcsname{%
		\noexpand\DOTSI
		\expandafter\noexpand\csname\next op\endcsname
		\noexpand\ilimits@
	}%
}
\makeatother
\shorttitle{Identifying Halos Using CWT}
\shortauthors{Li, Wang \& He}
\begin{document}
\begin{CJK*}{UTF8}{gbsn}
\title{CWTHF: Identifying Dark Matter Halos with Continuous Wavelet Transform}

\author[0009-0003-1625-8647]{Minxing Li (李敏行)}
\affiliation{College of Physics, Jilin University, Changchun 130012, P.R. China.}

\author[0000-0003-4064-417X]{Yun Wang (王云)}
\affiliation{College of Physics, Jilin University, Changchun 130012, P.R. China.}

\author[0000-0001-7767-6154]{Ping He (何平)}
\affiliation{College of Physics, Jilin University, Changchun 130012, P.R. China.}
\affiliation{Center for High Energy Physics, Peking University, Beijing 100871, P.R. China.}

\correspondingauthor{Ping He}
\email{hep@jlu.edu.cn}

\begin{abstract}
Cosmological simulations are an important method for investigating the evolution of the Universe. In order to gain further insight into the processes of structure formation, it is necessary to identify isolated bound objects within the simulations, namely, the dark matter halos. The continuous wavelet transform (CWT) is an effective tool used as a halo finder due to its ability to extract clustering information from the input data. In this study, we introduce CWTHF (Continuous Wavelet Transform Halo Finder), the first wavelet-based, MPI-parallelized halo finder, marking a novel approach in the field of cosmology. We calculate the CWT from the cloud-in-cell (CIC) grid and segment the grid based on the local CWT maxima. We then investigate the effects of the parameters that influence our program and identify the default settings. A comparison with the conventional friends-of-friends (FOF) method demonstrates the viability of CWT for halo finding. Although the actual performance is not faster than FOF, the linear time complexity of $\mathcal{O}(N)$ of our identification scheme indicates its significant potential for future optimization and application.
\end{abstract}


\section{Introduction}
\label{sec:intro}

Cosmology simulations have grown increasingly vital in the study of structure formation since the concept was first introduced \citep{Peebles1974, Press1974}. These simulations provide a powerful tool for understanding the complex processes involved in the evolution of the universe. By modeling the interactions between dark matter, baryonic matter, and radiation, cosmologists can gain insights into how galaxies, galaxy clusters, and other large-scale structures formed over time. As computational power has advanced, these simulations have become more sophisticated, allowing researchers to model larger volumes of the universe with higher resolution \citep{Davis1985}. This has led to a better understanding of the initial conditions and the physical processes that drive structure formation. Furthermore, simulations help in testing and refining theoretical models by comparing their predictions with observational data, thus contributing significantly to our knowledge of the universe's structure and evolution \citep{Frenk2012, Kuhlen2012}.

In such simulations, the density of the universe is typically represented by particles, including dark matter, gas, and stars. To extract structural information from these particles, it is essential to identify isolated, gravitationally bound objects, namely dark matter halos. In recent years, the rapid advancement of computer science has enabled researchers to conduct simulations with billions or even trillions of particles \citep[e.g.,][]{Dave2019,Nelson2019,Villaescusa2020}. Although some researchers can utilize these data without identifying isolated halos \citep[e.g.,][]{Balaudo2024}, the increasing volume of data necessitates the development of faster algorithms for data processing. Furthermore, while it is commonly stated that a dark matter halo is a `gravitationally bound object', this alone is insufficient to definitively identify halos from simulation datasets \citep{Knebe2013}. The precise definition of a dark matter halo remains a contentious issue. The lack of a universally accepted definition, coupled with the need to analyze simulation data, has driven the development of halo-finding algorithms. Below, we introduce some of them roughly according to the timeline.

The first is the spherical-overdensity (SO) halo finder \citep{Press1974}, which selects each particle and calculates the radius within which the density reaches a specified threshold. If a particle lies within the radius of another particle whose radius is larger than this particle, this particle will be discarded. The positions and radii of the remaining particles form the final catalog of `condensations'. Despite its $\mathcal{O}(N^2)$ time complexity and the assumption of spherical symmetry, this method effectively identifies dense groups from complex particle data. \citet{Lacey1994} had optimized the SO to avoid $\mathcal{O}(N^2)$ time complexity. The second is the friends-of-friends (FOF) halo finder \citep{Davis1985}, which links particles whose distances are below a certain fraction of the average particle distance. Particles linked together form groups, and groups with too few particles are discarded. Although the FOF method does not assume spherical symmetry, it may incorrectly link adjacent structures, leading to the so-called `linking bridge' problem \citep{Springel2001}. These two pioneers represent two fundamental approaches to halo finding: the SO-based density peak locators and the FOF-based particle collectors \citep{Knebe2011, Knebe2013}. The former identifies density maxima and gathers particles around these peaks, while the latter connects particles based on specific criteria. It can be argued that modern halo finders incorporate at least one of these two methods.

The DENMAX halo finder calculates the grid density and density gradients using the triangular-shaped cloud (TSC) method \citep{Bertschinger1991, Gelb1994, Goetz1998}. It then moves all particles in the direction of the density gradient until they reach local maxima. Particles that converge at the same density peak form a group. This algorithm is physically analogous to a highly viscous, perfectly compressible fluid flowing in a fixed gravitational field. The concept of moving particles toward density peaks was later adopted by the Hierarchical Object Partitioning (HOP) algorithm, which moves particles based on smoothed particle density rather than grid density \citep{Eisenstein1998}. Similarly, Spline Kernel Interpolative DENMAX (SKID), a successor of DENMAX, also utilizes smoothed particle density, which is suitable for Lagrangian $N$-body simulation \citep{Weinberg1997, Stadel2001}.

The Amiga's Halo Finder (AHF) employs a refined grid to identify cosmic structures \citep{Knollmann2009}. AHF calculates the TSC grid using particle data on a coarse grid. If the particle density in a given grid exceeds a specified threshold, the grid is refined to a finer resolution, and the particles within it are reassigned to the refined grid. The halo catalog is constructed based on isolated regions at each level of the grid hierarchy. Similarly, the Adaptive Spherical Overdensity Halo Finder (ASOHF) also uses an adaptive mesh refinement scheme \citep{Planelles2010, Valles2022}. However, unlike AHF, which assigns particles to their corresponding isolated regions, ASOHF applies the SO method independently at each level.

The Robust Overdensity Calculation using K-Space Topologically Adaptive Refinement (ROCKSTAR) initially congregates particles into FOF groups using a very large linking length. Subsequently, it divides these FOF groups into a hierarchical structure of subgroups. This is achieved by adaptively reducing the linking length in phase space until the subgroups contain a specific fraction of particles from their parent halo \citep{Behroozi2013}. The halo catalog is then constructed based on seed halos, which are derived from the subgroups without any further sublevels.

The Competitive Assignment to Spherical Overdensities (COMPASO) identifies structures at three levels \citep{Hadzhiyska2022}. At the largest level, the algorithm employs a modified FOF algorithm to link particles that exhibit a sufficiently high local density. At the subsequent two levels, the SO method is applied, centered on the densest particle. Particles located outside the SO radius, at a distance 0.8 times the radius, are considered as potential halo candidates. Among these candidates, the densest particle is identified. The SO method is then iteratively applied to the entire FOF group until no further dense particles remain among the eligible candidates. For these SO nuclei (the densest particles), particles are competitively assigned based on their enclosed density relative to these nuclei.

The halo finders, such as HOP and SKID, are designed using the kernel density estimation (KDE) method. While the KDE method is capable of deriving a continuous density distribution from discrete particle data \citep{Li2018}, the wavelet method, by contrast, not only can obtain a continuous density distribution from discrete particle data, but also offers a distinct halo boundary. In a previous study, we employed the continuous wavelet transform (CWT) technique to identify halos from a pseudo-two-dimensional (2D) dataset, which was derived from three-dimensional (3D) simulation data through a process of dimensional reduction \citep{Li2024}. Our work has demonstrated the applicability of the CWT in identifying halos from a 2D dataset. The wavelet transform technique has become a pivotal tool in cosmology, enabling the analysis of cosmic structure clustering across both spatial and scale domains \citep[e.g.,][]{Freeman2002, Patrikeev2006, Ciprini2007, Romeo2008, Wang2008, Gonzalez2010, Djafer2012, Vavilova2018,Wang2022a,Wang2022b}. The structures present in the given data will result in a high CWT value at the location of these structures, and the size of these structures can be determined by the scale of the cross-scale maximum \citep{Bendjoya1991, Bijaoui1992, Slezak1993, Grebenev1995, Flin2006}. Furthermore, the precise configuration of these structures can be ensured by segmenting the CWT from local maxima \citep{Flin2006, Baluev2020}. 

\input{flowchart.tex}

\begin{figure}[t]
    \centering
    \includegraphics[width=0.4\textwidth]{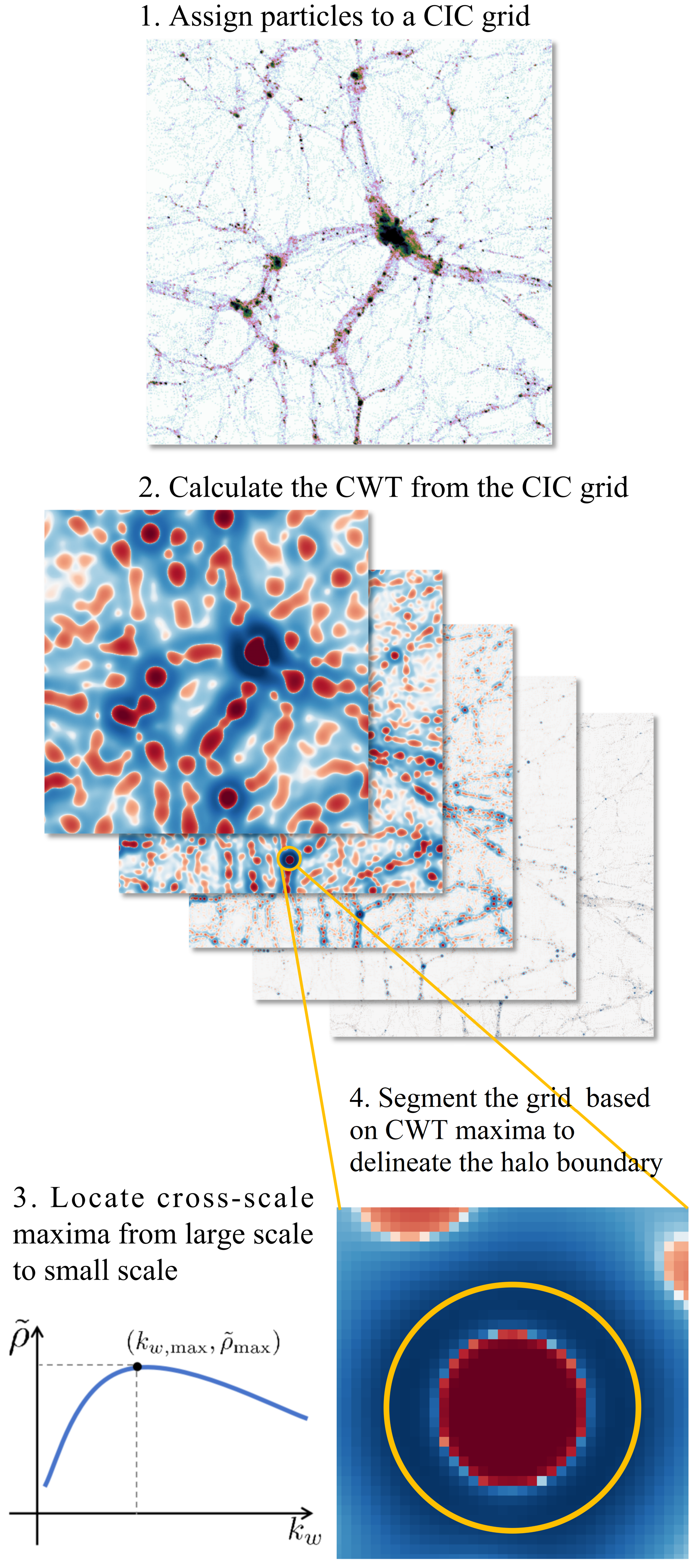}
    \caption{The visualized CWTHF algorithm and the delineation of the halo boundary in the 4D wavelet space.}    
    \label{visual_CWT}
\end{figure}

The current paper presents an updated and optimized version of our previous 2D algorithm to the real 3D case, introducing the continuous wavelet transform halo finder (CWTHF). Unlike traditional methods that directly identify structures in the density field, the CWTHF employs a multi-scale analysis to identify halos in the 4D (3D physical plus 1D scale) wavelet space. This methodology allows for the selection of specific scales for investigation, thereby enabling the identification of large halos with minimal resources in a time-efficient manner. The multiscale technique is a powerful tool in cosmology and has been used to detect various cosmological features \citep{Aragon2007}. It is important to note that any halo finder inherently has its own definition of a halo. In the case of CWTHF, the position and extent of the halos are contingent upon the location of the cross-scale maxima and the declining CWT value.

In this paper, we provide a comprehensive description of the CWTHF, including its parameters and parallel implementation strategy. We visualize the identification results obtained using the CWTHF and compare them with those from the FOF method. This paper is structured as follows. In Section~\ref{sec:methodology}, we introduce the CWTHF in detail. Section~\ref{sec:result} presents the identification results of the CWTHF and a comparison with the FOF catalog. Section~\ref{sec:parameters} investigates the effects of parameters on the CWTHF and provides a set of default values. Finally, Section~\ref{sec:concl} summarizes the conclusions regarding the CWTHF.

Incidentally, we have also developed a series of wavelet-based analysis techniques for various cosmological studies. For details of these CWT techniques, please refer to \citet{Wang2021, Wang2022a, Wang2022b, Wang2023, Wang2024a, Wang2024b, Wang2024c, Wang2025a}.

\section{The Methodology of CWTHF}
\label{sec:methodology}

We provide a detailed introduction to the CWTHF methodology. The halo identification algorithm in each iteration consists of the following steps.
\begin{enumerate}
  \item Select the dimensionless scale parameter $k_w$ in this iteration and calculate $N_g$ with $k_w$\footnote{As in \citet{Li2024}, $k_w \equiv w(L_{\rm box}/1000)$, where $L_{\rm box}$ represents the length of the simulation box, and $N_g$ is specified in Figure~\ref{approach_flowchart}.}.
  \item Assign all the particles to a $N_g^3$ cubic grid with Cloud-in-Cell (CIC) scheme.
  \item Calculate the grid CWT at the aforementioned scale. 
  \item Identify local maxima in the grid CWT and correct the bias caused by finite grid resolution.
  \item Segment the grid based on these CWT maxima and calculate the grid volume of these structures.
  \item Assign the particles without halo ID to the segmented grid, particles belonging to the same structure are assigned the same temp ID and form a group.
  \item Dispose of groups whose density is less than 4$\overline{\rho}$, where $\overline{\rho}$ is the average density in the entire simulation box.
  \item Dispose of groups that do not contain a cross-scale maximum.
  \item Conduct a self-boundness check on all the remaining groups.
  \item Preserve the particles that pass the self-boundness check and assign the halo ID.
\end{enumerate}

We present the flowchart and a visualized description of the CWTHF in Figures~\ref{approach_flowchart} and \ref{visual_CWT}. Two arrays, \texttt{calculating\_grid} and \texttt{current\_grid}, are utilized to store the grid CWT for the scale currently being processed, as well as for a smaller scale. In the initial iteration, the program only calculates the grid CWT and updates the aforementioned two arrays. In subsequent iterations, the grid CWT of the largest scale is accessed. By analogy, in the $n$-th iteration, the program deals with the CWT from the ($n-1$)-th largest scale. These two arrays are used for cross-scale comparison: for each local maximum in the \texttt{calculating\_grid}, its physical position is identified and evaluated to ascertain whether it exceeds the values of all grid points near the same physical position on an adjacent smaller scale, that is, the \texttt{current\_grid}. This process allows us to confirm that any maximum in 3D physical space is also a maximum in 4D space, which includes the 3D physical dimensions plus one additional scale dimension.

\subsection{Calculation of Grid CWT}
\label{sec:fastCWT}

In the 2D test \citep{Li2024}, the CWT is calculated directly based on the particle distribution and then sampled onto a regular grid. The grid CWT calculated by this approach is accurate but comes at the cost of increased computational costs. Indeed, the computational costs of the CWT approach are approximately two orders of magnitude higher than those of the FOF method, even in the simple 2D case. Thus, a fast algorithm is necessary to extend the CWT method to the case of 3D space. The fast algorithm in the current work for calculating the grid CWT can be summarized in two steps. The first step is to assign all the particles to a grid using a suitable window function, such as the CIC scheme. The CIC grid can be considered an approximation of the original particle distribution. In other words, the particles are relocated to the grid points, and the pseudo-particles situated at these grid points $\mathbf{x}_{g}$ are used to reconstruct the particle density field.
\begin{flalign}
    \sum_{i}m_i\delta^{(\rm D)}(\mathbf{x}-\mathbf{x}_i) \approx \sum_{\mathbf{g}}m_{\mathbf{g}}\delta^{(\rm D)}(\mathbf{x}-\mathbf{x}_{\mathbf{g}}),
\end{flalign}
where $\mathbf{g}$ denotes the vector index of grid points, and $\mathbf{x}_{\mathbf{g}}$ and $m_\mathbf{g}$ represent the position and the CIC mass of grid point $\mathbf{g}$, respectively. The second step is to calculate the grid CWT using the CIC grid:
\begin{flalign}
\label{CWT_grid}
  \tilde{\rho} (w,\mathbf{x}) = &\int_{-\infty}^{\infty}\rho (\mathbf{u})\Psi (w,\mathbf{x}-\mathbf{u})\dd\mathbf{u} \nonumber \\
  = & \int_{-\infty}^{\infty}\sum_{i}m_i\delta^{(\rm D)}(\mathbf{u}-\mathbf{x}_i)\Psi (w,\mathbf{x}-\mathbf{u})\dd\mathbf{u}  \nonumber \\
  \approx & \int_{-\infty}^{\infty} \sum_{\mathbf{g}}m_{\mathbf{g}}\delta^{(\rm D)}(\mathbf{u}-\mathbf{x}_{\mathbf{g}})\Psi (w,\mathbf{x}-\mathbf{u})\dd\mathbf{u}  \nonumber \\
  = & \sum_{\mathbf{g}}\int_{-\infty}^{\infty} m_{\mathbf{g}}\delta^{(\rm D)}(\mathbf{u}-\mathbf{x}_{\mathbf{g}})\Psi (w,\mathbf{x}-\mathbf{u})\dd\mathbf{u}  \nonumber \\
  = & \sum_{\mathbf{g}}m_{\mathbf{g}}\Psi (w,\mathbf{x}-\mathbf{x}_{\mathbf{g}}),
\end{flalign}
where we compute the CWT $\tilde{f}(w,\mathbf{x})$ of a signal $f(\mathbf{x})$ as follows,
\begin{flalign}
    \tilde{f}(w,\mathbf{x}) =\int_{-\infty}^{\infty}f (\mathbf{u})\Psi (w,\mathbf{x}-\mathbf{u}) \dd\mathbf{u}.
\end{flalign}

\begin{figure}[t]
    \centering
    \includegraphics[width=0.44\textwidth]{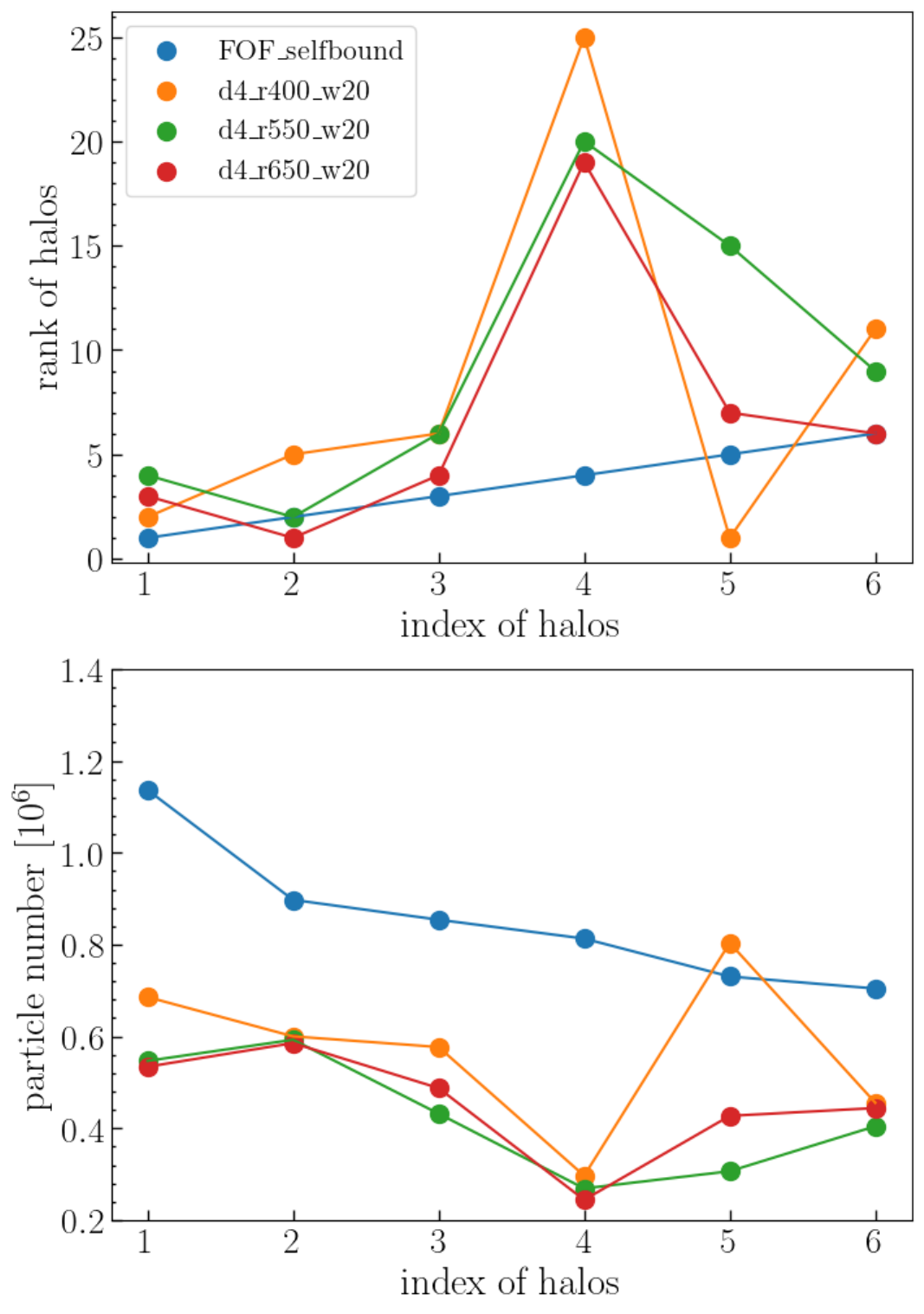}
    \caption{The rank (top) and the mass (bottom) of the six largest halos in the FOF catalog, as observed in other uncorrected catalogs. For the meaning of the catalog names see Section~\ref{sec:result}.}
    \label{unstable}
\end{figure}

\begin{figure}[t]
    \centering
    \includegraphics[width=0.44\textwidth]{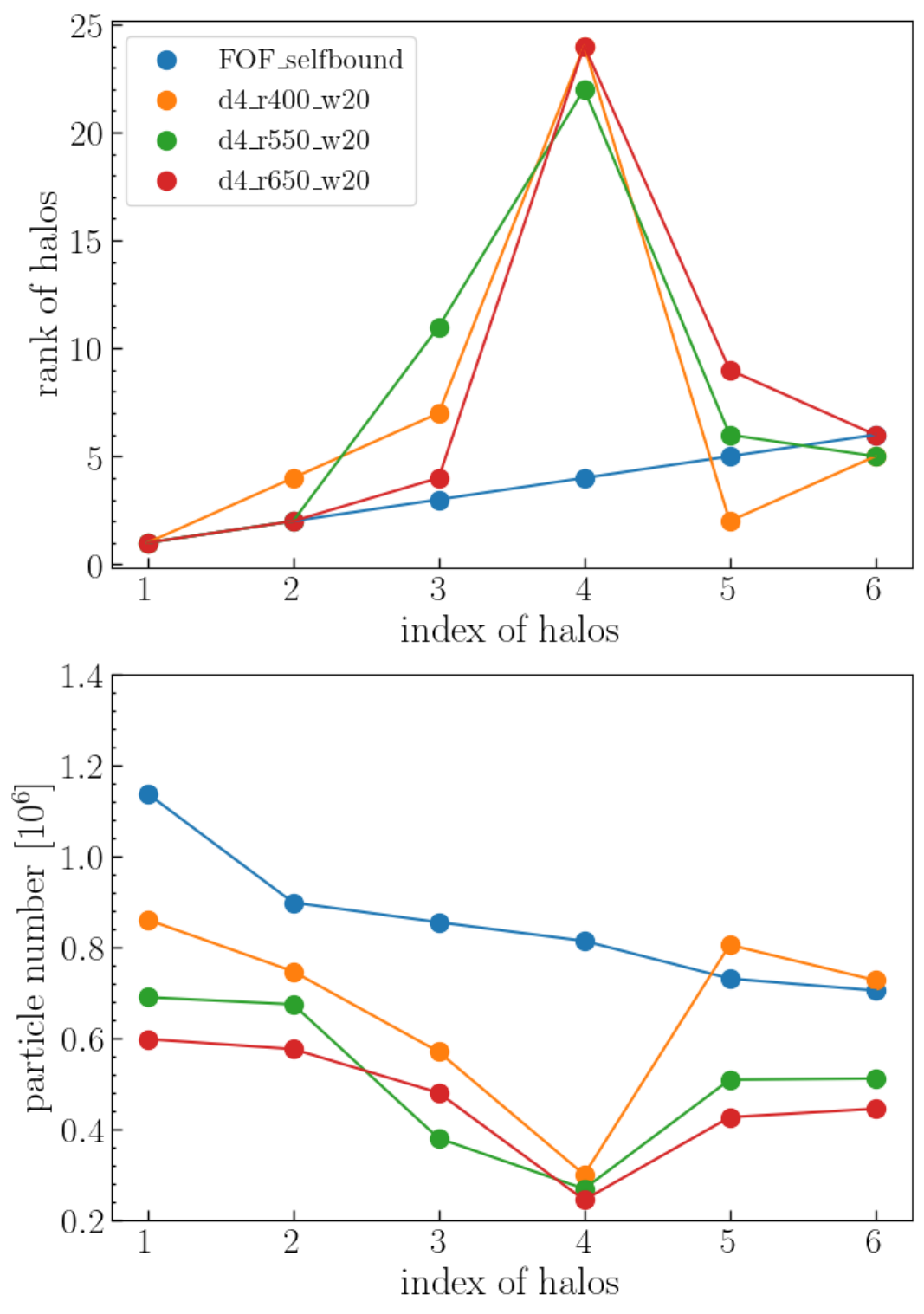}
    \caption{The rank (top) and the mass (bottom) of the six largest halos in the FOF catalog, as observed in other corrected catalogs.}
    \label{stable}
\end{figure}

The calculation of the grid CWT $\tilde{\rho} (w,\mathbf{x})$ is simple because the positions of these pseudo-particles are regular. First, compute the wavelet $\Psi (w,\mathbf{x})$ on a regular grid, where the grid spacing is set to match the CIC grid mentioned earlier. Then, for each grid point $\mathbf{g}$ with $m_{\mathbf{g}}>0$, weight the wavelet by $m_{\mathbf{g}}$ and add the weighted wavelet to the grid points centered around $\mathbf{x}_{\mathbf{g}}$. This approach effectively transforms the calculation of the grid CWT into a process of addition of Numpy.ndarray, which is significantly more time-efficient than the algorithm presented in the previous work.

Essentially, the algorithm uses a discretized integral definition to perform the convolution calculation, which results in a linear time complexity. Each actual particle can be represented by up to eight pseudo-particles within the grid. Consequently, the total number of required calculation operations is limited to $8 \times N_{p} \times N_w$, where $N_p$ is the number of particles and $N_w$ is the number of grid points used to sample the wavelet.

In the present study, the isotropic Mexican hat (MH) wavelet is employed in 3D space, as 
\begin{flalign}
    \Psi (w,\mathbf{x}) = w^{3/2}(6 - w^2 r^2)e^{-\frac{w^2 r^2}{4}},
\end{flalign}
which is derived from the first derivative of the Gaussian function with respect to $w$ \citep{Wang2021, Wang2022a}. The use of the isotropic wavelet helps to streamline the calculation process while preserving the ability to detect anisotropic structures \citep{Vavilova2018}.

\subsection{Correction of Peak Grid Points}
\label{sec:gridbias}

In order to reduce computational costs, the CWT is calculated using the grid density, with halos identified based on local maxima in the grid. However, it should be noted that the CWT values at these grid points do not represent the so-called `true' maxima. Instead, the value at a specific grid point represents the average value within the corresponding grid cube. Therefore, using the maxima grid points to identify peaks in 4D space introduces two inherent biases. First, because the grid is fixed at each scale, the position of the grid cube significantly affects the value of the grid maximum. Grid cubes that are closer to the `true' maxima have higher peak values. Second, using a smaller grid cube results in a higher peak value, as the averaging process excludes the outer, lower values. This causes halos to appear smaller because they are selected at a smaller scale with a larger CWT value.

Given the scale-specific characteristics of wavelets, a dynamic grid resolution is used to balance precision and computational efficiency. Reallocating the grid introduces bias in peak value calculations due to the two factors mentioned earlier. This bias leads to significant variability in halo sizes when the spatial resolution changes, as shown in Figure~\ref{unstable}. Moreover, the ranking of the largest halos varies randomly across different resolutions and does not match the FOF ranking.

In order to mitigate the impact of this phenomenon, we select six adjacent grid points to the maximum grid point, fit the maximum to a quadratic function in each dimension, and calculate the shift to the unbiased maximum. This approach effectively suppresses the bias introduced by the location of the grid cube. Regarding the second factor, we assume that the behavior of the CWT near the peak can be approximated by a 3D isotropic quadratic function and calculate its peak value for a smaller grid length (i.e., the grid length of the smaller scale). The details of these two fitting approaches are presented in Appendix~\ref{sec:correction}.

After correcting the peaks, as illustrated in Figure~\ref{stable}, the size and rank of these largest CWT halos stabilize, showing a trend where the size of the halos decreases with increasing grid resolution.

\begin{figure}[t]
    \centering
    \includegraphics[width=0.44\textwidth]{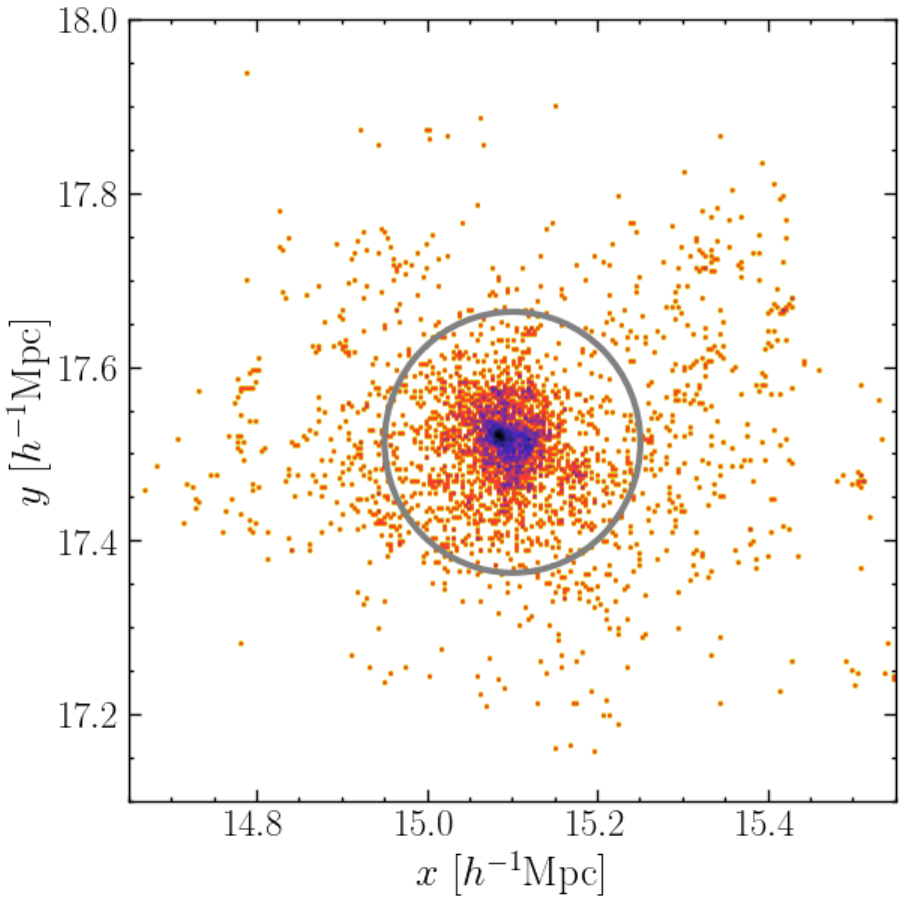}
    \caption{The scatter plot of a medium-sized CWT halo and its corresponding virial radius $r_{200}$ (gray circle). The color of the particles indicates their local density, with yellow representing low density and black representing high density. It is evident that the size of this halo is 2-3 times larger than $r_{200}$, resulting in a grid density that is 10-30 times smaller than 200$\overline{\rho}$.}
    \label{density_vir}
\end{figure}

\subsection{Grid Segmentation and Halo Boundary Delineation}

As observed in our previous work, any halo in the grid CWT exhibits a similar pattern: a positive peak surrounded by a negative ring \citep{Li2024}. This pattern arises from the alignment between the density peak and the positive peak, negative ring of the wavelet. We consider all regions surrounding the positive peak with CWT $>0$ as the dark halo itself, while regions with CWT $\leq 0$ are regarded as the exterior of the dark halo. In this way, we can clearly delineate the halo boundary.

In many previous studies, the extent of structures identified using the continuous or discrete wavelet transform is delineated according to certain threshold values \citep{Freeman2002, Mertens2015, Baluev2020}. In the context of halo identification, our previous work shows that defining the halo boundary based solely on positional data results in an overly sharp boundary. This is due to the inherent limitations of finite grid resolution. Fortunately, the gravity-bounded nature of dark matter halos allows for a self-boundness check with velocity information \citep{Stadel2001, Knollmann2009, Behroozi2013, Valles2022}.

To achieve a smoother boundary, the CWT halo boundary should not merely be defined by CWT $=0$, but should instead be extended outward to the lowest point (valley) of the CWT. However, extending the CWT boundary in this manner also include unrelated particles into the halos. Therefore, a self-boundness check is used to determine whether a particle truly belongs to a halo.

The self-boundness check involves the following steps: First, locate the center of mass as the halo center. Second, assuming spherical symmetry, calculate the gravitational potential and, subsequently, the local escape velocity for each particle. Third, exclude particles whose velocities exceed the local escape velocity and update the halo center position based on the remaining particles. Repeat these steps until no particles exceed the local escape velocity. If more than 40\% of the particles are excluded, the entire group is considered unbound and is discarded. The details of the self-boundness check are similar to those used in the AHF method and can be found in the appendix of \citet{Knollmann2009}.

\subsection{Cross-scale Verification and Density Threshold}

A cross-scale verification is necessary to determine which maxima correspond to actual halos rather than Poisson noise, and the scale at which a maximum is identified indicates the size of these structures \citep[e.g.][]{Bendjoya1991, Slezak1993, Grebenev1995, Lazzati1999, Cayon2000, Patrikeev2006, Flin2006, Hayn2012}.

A cross-scale verification of all the maxima is a time-consuming process, especially since Poisson noise inherent in particle data also produces a significant number of maxima. To address this issue, we employed a Monte Carlo simulation in our previous work. However, running a Monte Carlo simulation requires a significant amount of computational resources, and the threshold obtained through this method does not have a clear physical interpretation. Therefore, we propose a thresholding method based on the physical criterion that each maximum should correspond to a structure with a density greater than $4\overline{\rho}$.

The threshold density is considerably lower than that predicted by the spherical collapse model, any value greater than 1 and less than 7 produces acceptable results (see parametric studies presented in Section~\ref{sec:parameters}). This discrepancy can be attributed to two factors. First, the program uses the grid volume instead of the virial volume to quantify the volume of these structures, as shown in Figure~\ref{density_vir}. In the case of low-to-medium mass structures, the grid volume is larger than the virial volume because it already includes the low-density external region. Second, regarding the density criterion, the goal is simply to filter out suspicious maxima, rather than to achieve the final identification result. During the density check phase, the structures input to the program are only a rough segmentation of these particles. Applying a strict criterion at this stage could potentially introduce bias into the final outcome.

\begin{figure}[t]
    \centering
    \includegraphics[width=0.44\textwidth]{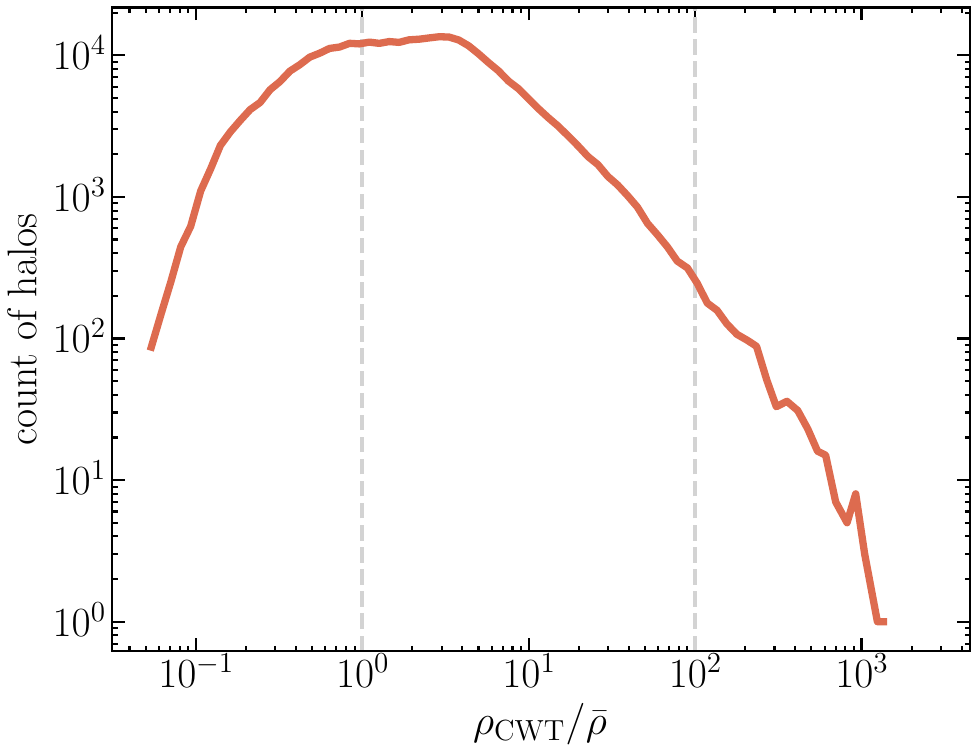}
    \caption{The halo density function of non-threshold CWT halos with two vertical lines denoting $\overline{\rho}$ and $100\overline{\rho}$, respectively. There is a significant number of low-density halos with densities even less than $\overline{\rho}$.}
    \label{smallrho}
\end{figure}

The halo density function of non-threshold CWT halos is shown in Figure~\ref{smallrho}. Two key pieces of information can be obtained from the density distribution. First, it is evident that the grid density of CWT halos is lower than the virial density predicted by the spherical collapse model. Setting a density threshold $\rho_{\rm th} \sim 100\overline{\rho}$ results in a sharp decline in the number of identified halos, with only a few thousand remaining. In contrast, the FOF method identifies two orders of magnitude more halos. Second, using only a cross-scale maximum check, our CWT method identifies approximately one-third of the halos with densities lower than $\overline{\rho}$, suggesting that these halos may not be physically realistic.

\subsection{Parallelization}
\label{sec:Parallelization}

Analyzing a dataset containing a large number of particles requires the use of parallelization, due to the limitations of a single-core CPU. The parallelization of the CWTHF is implemented using the Message Passing Interface (MPI), which is integrated into the Python environment through MPI4py.

During the entire calculation process of the CWTHF, the computation of the grid CWT and the segmentation of the grid CWT account for more than 95\% of the total computation time. Therefore, the parallelization strategy should primarily target these two components, although the remaining elements are also amenable to parallelization. Additionally, the use of the CIC grid facilitates the application of the domain decomposition technique.

In the calculation of the grid CWT, the entire CIC grid is divided into equal-sized sections along the z-axis wherever feasible. These sections are then assigned to different CPUs, and the grid CWT is calculated independently for each section. As stated in Section~\ref{sec:fastCWT}, the CIC grid represents a regularized particle distribution. Consequently, the calculation of the CWT between different grid points is inherently independent, and the final result is simply the superposition of the CWT contributions from disparate grid points. The CWT values computed by distinct CPUs are aggregated at the appropriate positions in the entire grid, thereby enabling the parallelization of the CWT calculation through MPI.

In the case of a grid segmentation, the grid CWT is still divided along the z-axis, but with extended ghost cells\footnote{In computational fluid dynamics and numerical simulations, ghost cells refer to virtual cells used in the computational grid to handle boundary conditions. These cells are located outside the computational domain, but can implicitly express the boundary conditions of the physical surface through appropriate assignments.}. The process begins with identifying each maximum within the CWT sections. However, the delineation of the halo boundary is not entirely localized. Instead, the search for the CWT valley can extend up to a halo radius beyond the CWT section. Furthermore, in regions of high density, delineating the precise boundary requires knowledge of adjacent halos. Consequently, the CWT information from neighboring sections is also transmitted to these CPUs. The actual performance and parallelization efficiency of our program are presented in Section~\ref{sec:performance}.

\begin{figure*}[t]
    \centering
    \includegraphics[width=0.98\textwidth]{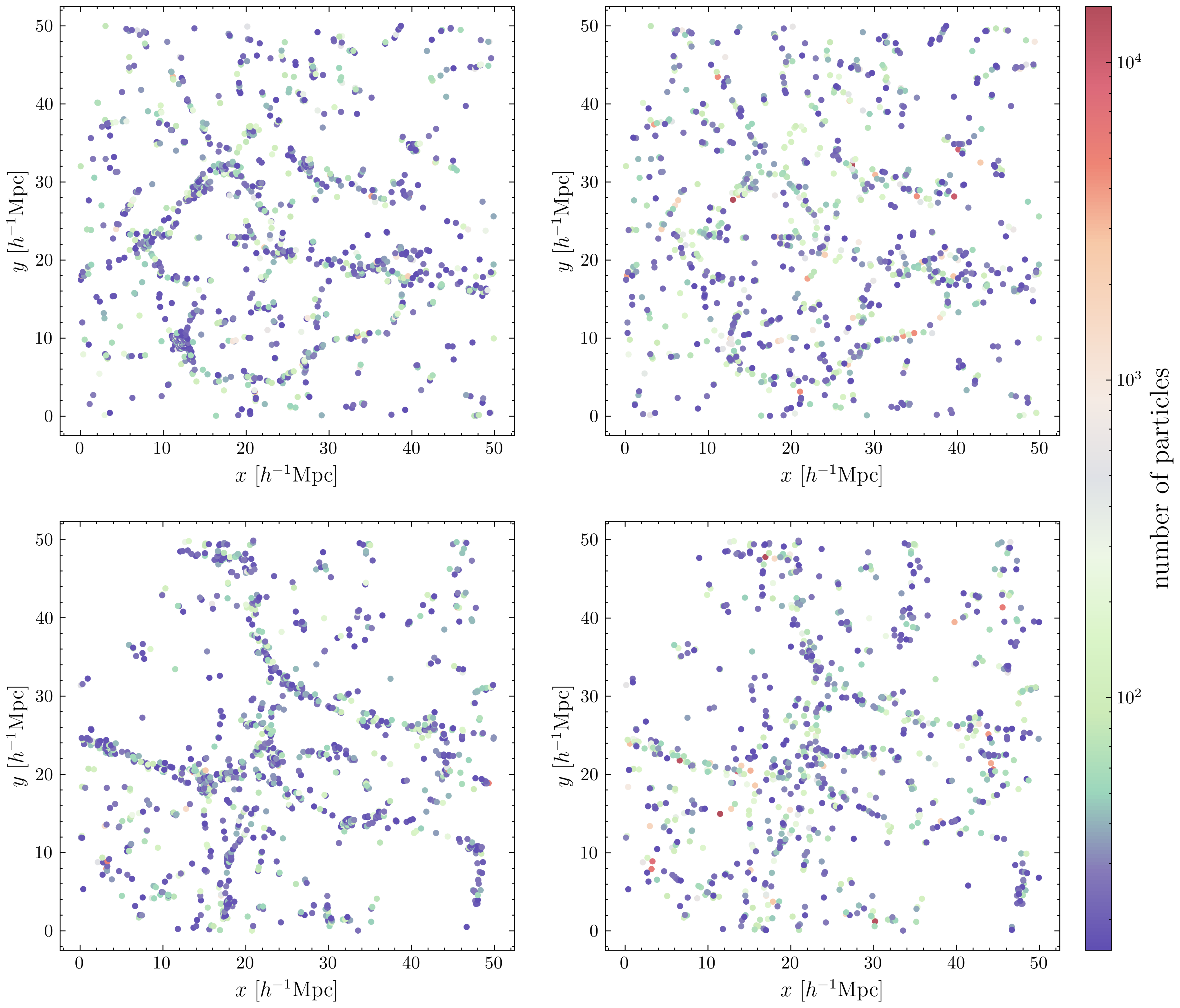}
    \caption{Left: The distribution of FOF halos in two slices located at ${\rm z} = 20.0\mpch$ and $43.2\mpch$, each with a thickness of $\Delta{\rm z} = 0.3\mpch$. Right: The distribution of CWT halos in the same two slices.}
    \label{halo_distribution}
\end{figure*}

\section{Results}
\label{sec:result}

The tests of the CWTHF are conducted using the m50n512 dark matter sub-simulation from the SIMBA project, which has a box length of 50$\hmpc$ and contains $512^3$ dark matter particles \citep{Dave2019}. SIMBA is a meshless finite mass hydrodynamic cosmological simulation based on the GIZMO code \citep{Hopkins2015} and the Planck cosmological parameters \citep{Planck2016}. To assess the reliability of the CWTHF, a FOF catalog is generated for the same dataset using the yt-project \citep{Turk2011}, which serves as a reference for the halo attributes in the dataset.

Figure~\ref{halo_distribution} depicts the distribution of our CWT halos and their corresponding FOF halos. The dimensionless scale parameter $k_w$ ranges from 0.1 to 2.5, corresponding to a physical scale of 0.5$\mpch$ to 0.02$\mpch$ during halo identification. This section focuses on the \texttt{d4\_r400\_w20} resolution run, which has a density threshold of 4$\overline{\rho}$, a grid resolution of 400 points per $k_w$, and a scale resolution of 20\footnote{Similar naming conventions also apply to the \texttt{d4\_r550\_w20} and \texttt{d4\_r650\_w20} catalogs used in Figures~\ref{unstable} and \ref{stable}.}. The distributions of the FOF and CWT halos are similar and form a web-like structure.

\begin{figure*}[t]
    \centering
    \includegraphics[width=0.98\textwidth]{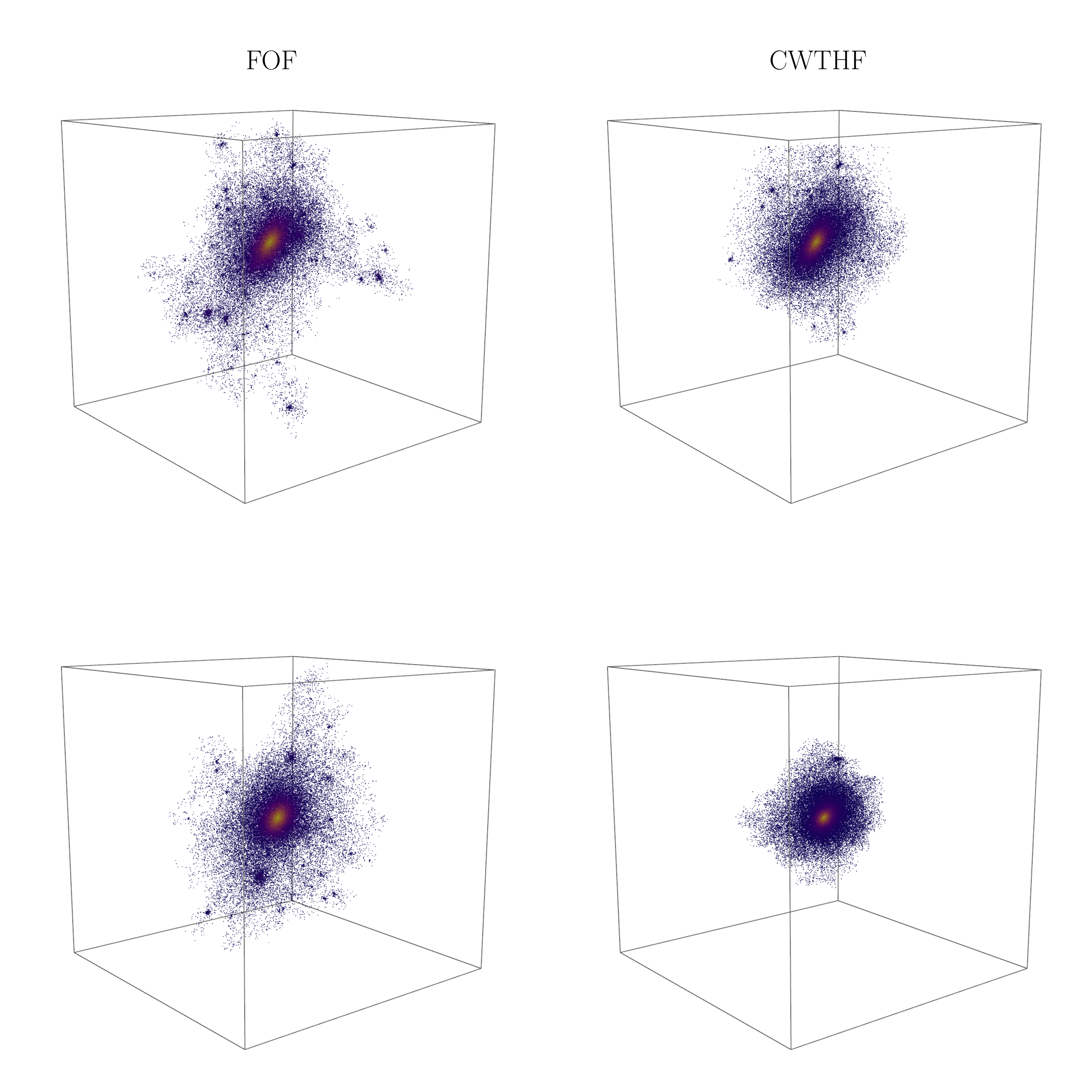}
    \caption{Left: The largest and the second largest FOF halos that survive the self-boundness check. Right: The corresponding CWT halos of the above two FOF halos. Both two halos exhibit an elliptical central structure.}
    \label{halo_compare}
\end{figure*}

The CWTHF identifies 162,872 CWT halos, with a corresponding number of FOF halos at 253,112. The primary difference between the CWT and the FOF catalog is that the CWT catalog has fewer small halos (represented by purple dots in the color map), which results in a more sparse distribution of CWT halos compared to the FOF halos. The observed discrepancy is mainly due to the self-boundness check process that is integral to the CWTHF. Unlike FOF halos, which are defined by the mere aggregation of closely spaced particles, all CWT halos are self-bounded. It is important to note that these FOF groups may not be dynamically stable; there is a high probability that they will become separated from each other, and as a result, these FOF groups may cease to exist within a few time steps.

The use of a self-boundness check in the FOF catalog has been shown to significantly reduce the discrepancy; the total number of FOF halos is reduced to 182,362, indicating that approximately 30\% of the FOF halos are unstable. However, the unbinding procedure used in the FOF catalog has been shown to result in an approximate 20\% decrease in the total number of particles within halos. The post-processing stage following the construction of the FOF catalog does not retain any particles. In the case of very large FOF halos, which inherit the `linking bridge' problem, the self-boundness check discards the entire set of particles without considering the possibility that a subset of particles may form smaller bounded structures.

Figure~\ref{halo_compare} demonstrates the two largest FOF halos and their corresponding CWT halos. The color of each particle corresponds to its density. These two halos from different catalogs have similar positions, masses, and even particle content, so they can be considered the same halos. As previously observed in the 2D test, the CWT halos are not larger than their FOF counterparts. Additionally, the boundary density of the CWT halos is observed to be slightly higher than that of the FOF halos, except in the largest CWT halo, where the boundary density is nearly identical to that of its FOF counterpart. 

Both CWT halos exhibit an elliptical central structure, confirming the ability of the spherical CWT to identify non-spherical structures. For these irregular halos, their high-density centers still align with the wavelets of similar size, thereby forming peaks in the 4D wavelet space. Furthermore, it is important to note that the information regarding the halo shape in the CIC density is not completely erased in the convolution process. The boundaries delineated in the CWT grid follow the shapes of the halos in the CIC grid, and the outward extension during the segmentation process further enriches the shapes of the CWT halos.

\begin{figure}[t]
    \centering
    \includegraphics[width=0.44\textwidth]{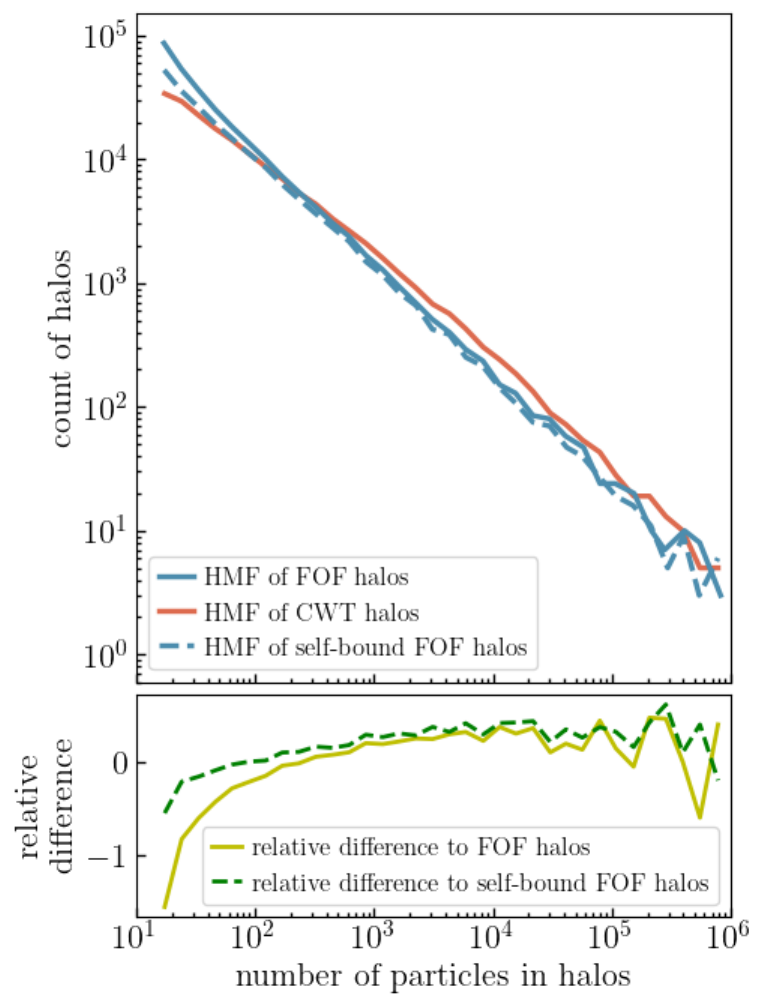}
    \caption{The HMFs of FOF and CWT halos (top) and the relative differences between the FOF and CWT catalogs (bottom). The solid red and blue lines represent the HMFs of CWT and FOF halos, respectively. The dashed blue line represents the HMF of FOF halos after an unbinding procedure.}
    \label{hmf}
\end{figure}

Compared to CWT halos, the configuration of large FOF halos is characterized by greater irregularity, manifesting as multiple branches that extend outward. It is reasonable to hypothesize that these branches are excluded because their characteristic scales (i.e., physical size) are much smaller than the main structure of the CWT halos. Eliminating the boundary threshold not only reduces the size discrepancy between FOF and CWT halos but also refines the boundaries of the CWT halos. With the extended outward boundary and the self-boundness check, the jagged shape of the halo boundary no longer exists, even if the grid resolution is lower than that of the low-resolution run in the 2D test.

The distribution of CWT and FOF halos in terms of halo mass can be further examined through the halo mass functions (HMFs), as shown in Figure~\ref{hmf}. The overall trends differ from those observed in the 2D test, with the most significant difference being that the halo number in the CWT catalog is lower than that in the FOF catalog, resulting in a much lower HMF value at the low-particle end. Additionally, neither the decline in the HMF in the mass range slightly above the minimum particle threshold nor the valley in the relative difference is observed. The unbinding procedure significantly reduces the discrepancy between the FOF and CWT HMFs at the low-particle end.

\begin{figure}[t]
    \centering
    \includegraphics[width=0.44\textwidth]{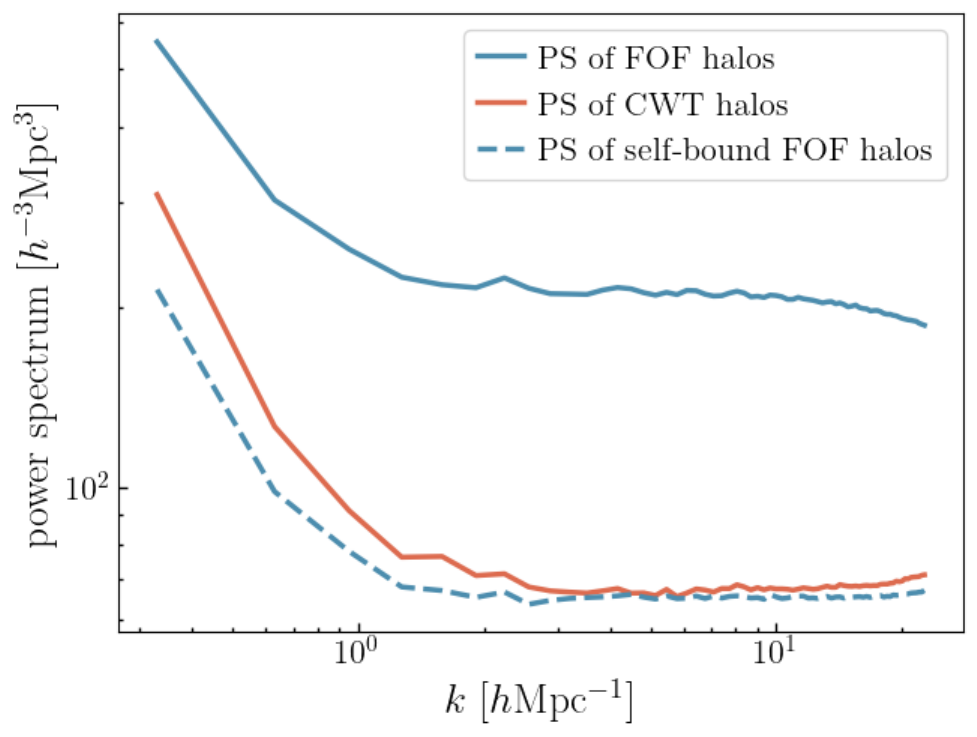}
    \caption{The power spectrum of FOF and CWT halos. The solid red and blue lines represent the power spectrum of CWT and FOF halos, respectively. The dashed blue line represents the power spectrum of FOF halos after an unbinding procedure. With the unbinding procedure, the shift between the FOF and CWT power spectra vanishes.}
    \label{ps}
\end{figure}

\begin{figure}[t]
    \centering
    \includegraphics[width=0.44\textwidth]{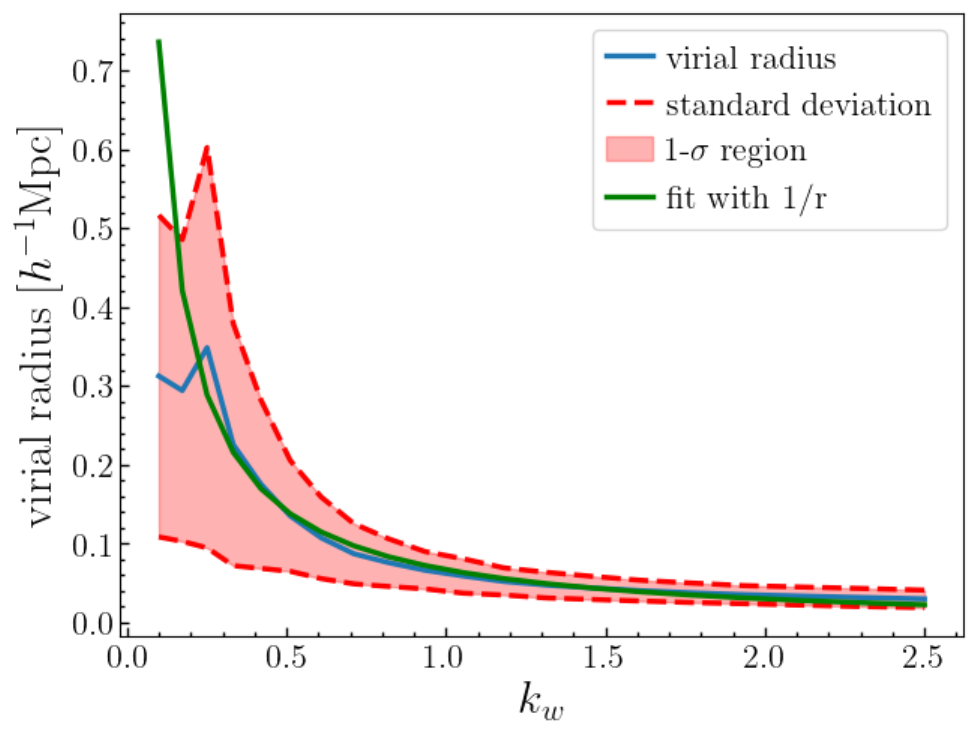}
    \caption{The relationship between the wavelet scales (dimensionless scale parameter) and the physical scales (virial radius) is illustrated by a solid blue line, with a 1-$\sigma$ uncertainty limit indicated by a dashed red line. The solid green line represents an inverse proportional function that has been fitted to this relationship.}
    \label{r_kw}
\end{figure}

\begin{deluxetable*}{lll}
\renewcommand{\arraystretch}{1.25}
\tablehead{\colhead{Name} & \colhead{Definition} & \colhead{Default Value} }
\tablecaption{Parameters in the CWTHF, their meanings, and the default values. \texttt{kw\_low} and \texttt{kw\_high} determine the lower and upper limits of the identification process. \texttt{w\_resolution}, \texttt{gdw\_resolution}, \texttt{n\_ref}, and \texttt{n\_min} are resolution parameters for the entire 4D wavelet space. \texttt{dens\_th} and \texttt{particle\_th} represent the minimum grid density and the minimum particle number for CWT halos, respectively. \texttt{ghost\_cell\_size} specifies the number of grid points that must extend outward from the divided sections when used in a multi-process environment.}
\tablenum{1}
\startdata
    \texttt{kw\_low}        & The lower limit of the dimensionless scale parameter $k_w$    & 0.1   \\ \hline
    \texttt{kw\_high}       & The upper limit of the dimensionless scale parameter $k_w$    & 2.5   \\ \hline
    \texttt{w\_resolution}  & The number of scales between \texttt{kw\_low} and \texttt{kw\_high}  &  Int((kw\_high - kw\_low)*8) \\ \hline
    \texttt{gdw\_resolution}  & The grid resolution of the GDW wavelet employed in CWTHF &  Int(1.5/$k_w$+5)  \\ \hline
    \texttt{n\_ref}         & The grid resolution of the CWT per $k_w$        & 400   \\ \hline
    \texttt{n\_min}         & The lower limit of the grid resolution          & 150   \\ \hline
    \texttt{dens\_th}       & The density threshold (grid density) of the structures after the segmentation process & 4 \\ \hline
    \texttt{particle\_th}    & The minimum particle number of halo        & 15   \\ \hline
    \texttt{ghost\_cell\_size} & The size of ghost cells when segmenting the CWT in a parallelized approach & 3\texttt{gdw\_resolution}\\ \hline
\enddata
\tablecomments{The default values of \texttt{kw\_low} and \texttt{kw\_high} are suitable only for simulations with a box length of 50$\mpch$ and $512^3$ particles.}
\label{tab:parameter}
\end{deluxetable*}

\begin{figure*}[t]
    \centering
    \includegraphics[width=0.98\textwidth]{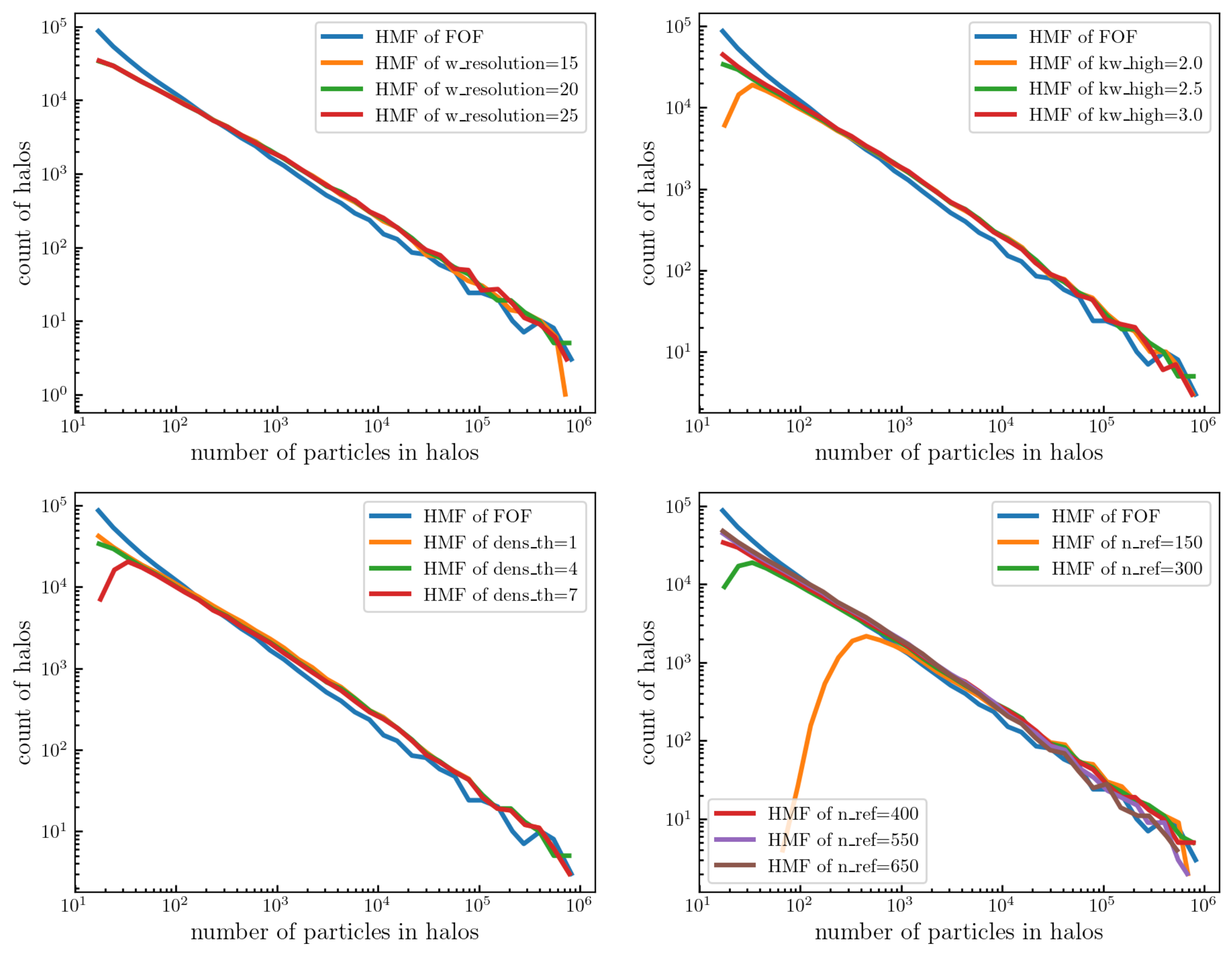}
    \caption{Effects of different parameters on the HMFs of CWT Catalogs. From the top left to the bottom right, we examine the effects of \texttt{w\_resolution}, \texttt{kw\_high}, \texttt{dens\_th}, and \texttt{n\_ref}, respectively. The solid blue line represents the HMF of the FOF catalog for comparison.}
    \label{HMF_para}
\end{figure*}

The number of FOF halos is 1.5 times greater than that of the CWT halos at the minimum particle threshold (15 particles). As the particle number increases, the number of CWT halos also increases, eventually reaching and then exceeding the number of FOF halos when the particle number reaches a few hundred. This excess stabilizes around 40\%. For the largest halos, their rarity introduces an observable scatter in the HMF; however, their overall numbers are generally comparable. As previously mentioned, the discrepancy between the FOF and CWTHF HMFs at the low-particle end is primarily due to the unbound halos identified by the FOF method.

In the context of HMFs for larger particle numbers, the excess in the number of CWT halos can be attributed to two main factors. First, the CWTHF excludes some outer structures of the largest halos and identifies them as separate, isolated halos. Second, the exclusive selection of unimodal structures further separates closely located structures, effectively addressing the so-called `linking bridge' problem. The result of these two processes is the division of large halos into multiple smaller halos, which increases the number of medium-sized halos and raises the HMF.

The halo power spectrum for the test dataset was calculated by assigning the halos to a $512^3$ grid using a weighted CIC method, with the results presented in Figure~\ref{ps}. The power spectra of both the FOF and CWT halos show a downward trend; however, the FOF power spectrum is shifted to higher amplitudes across all wavenumbers. This discrepancy may be due to the presence of a small number of unphysical large halos identified by the `linking bridge' problem inherent in the FOF method. By simply employing an unbinding procedure to remove unbound halos (some of which contain over two million particles), the power spectra of the remaining FOF halos exhibit amplitudes comparable to those observed for CWT halos.

To gain further insight into the physical counterpart of the CWT method, Figure~\ref{r_kw} illustrates the relationship between wavelet scales and physical scales. Given that $k_w \propto w \propto 1/a$, where $a$ is the physical scale of the wavelet, it is reasonable to assume that a similar relationship exists between the wavelet scale and the halo size at that scale. The fitting results yield a function of the form
\begin{flalign}
    r_{\rm vir} = \frac{0.091}{k_w} - 0.012,
\end{flalign}
in which $r_{\rm vir}$ is in units of $\mpch$. Considering the results mentioned earlier, an approximate relation can be derived using the definition of $k_w$, as
\begin{flalign}
    r_{\rm vir} \simeq \frac{1.8}{w}.
    \label{relation}
\end{flalign}
It should be noted that in the curve fitting process, halos from the three largest scales are excluded. This is because it is not feasible for halos to achieve a very large virial radius within the context of the CDM model. In this range, increasing the search scale does not lead to larger halos, and the simple $1/r$ relationship is no longer applicable.

\begin{figure*}[t]
    \centering
    \includegraphics[width=0.98\textwidth]{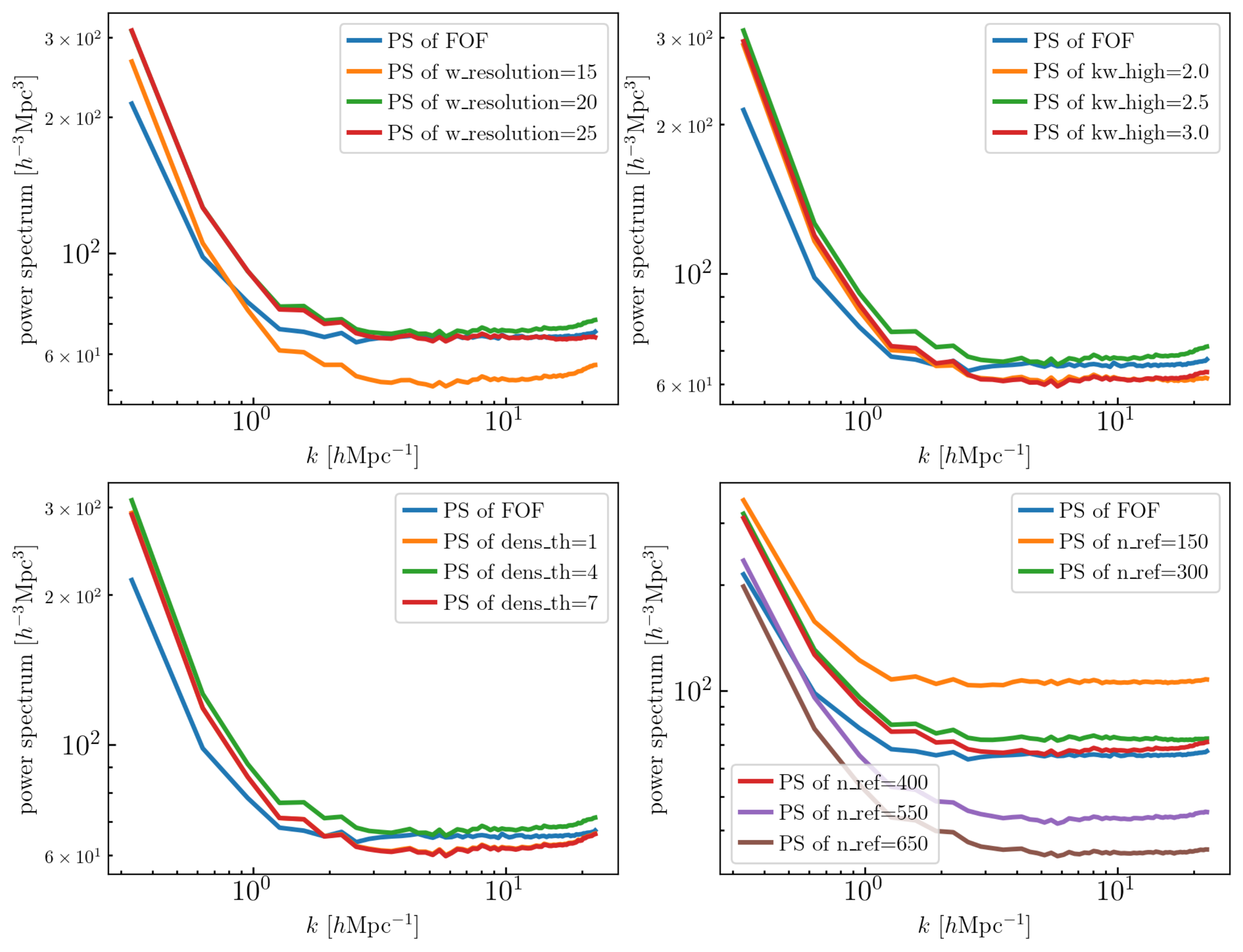}
    \caption{Similar to Figure~\ref{HMF_para}, but examining the effects on the HPS. A self-boundness check is applied to the FOF HPS to reduce the shift between the FOF and CWT HPS.}
    \label{PS_para}
\end{figure*}

\section{Parameters in CWTHF}
\label{sec:parameters}

In this section, we examine the effects of parameters on the CWTHF. The parameter tests are based on two statistical properties of the halos: the HMF and the halo power spectrum (HPS). The HPS is calculated by assigning all halos to a CIC grid and weighting them by their mass (particle number). The standard is set as the \texttt{d4\_r400\_w20} catalog, which has parameters \texttt{w\_resolution}=20, \texttt{kw\_high}=2.5, \texttt{dens\_th}=4, and \texttt{n\_ref=400}. The parameters used in the CWTHF are summarized in Table~\ref{tab:parameter}.

As shown in Figure~\ref{HMF_para}, the HMF of the CWT catalog remains stable above approximately 500 particles, except in cases where the spatial resolution is very low (\texttt{n\_ref} = 150). Within this range, all four sets of HMF are nearly identical and higher than the HMF of the FOF. However, as the halo mass exceeds several hundred thousand particles, variations begin to appear within these sets of HMF.

Unlike other parameters, the impact of \texttt{w\_resolution} on the HMF is significant only in the large mass range (see the top left panel of Figure~\ref{HMF_para}). Changing \texttt{w\_resolution} affects the sets of $k_w$ used by the CWTHF to calculate the grid CWT, which in turn influences the locations of maxima in the 4D wavelet space and their CWT boundaries. This leads to changes in the sizes of large CWT halos. However, for small halos, their CWT boundaries already encompass the no-particle voids, ensuring that their particle content is not affected by this factor.

Unlike \texttt{w\_resolution}, \texttt{dens\_th} affects the HMF primarily at the low-mass end, causing the HMF to decrease at high-density thresholds (see the bottom left panel of Figure~\ref{HMF_para}). This occurs because the density of large halos is not affected by the difference between the grid density and the virial density. Even if the grid density underestimates the halo density, the additional grid resolution indicated by \texttt{n\_min} on a large scale mitigates this issue. The underestimation is more pronounced for smaller halos due to their broader boundaries. For a spatial resolution of \texttt{n\_ref} around 400, the corresponding density threshold should be set at 4 or lower.

The parameters \texttt{kw\_high} and \texttt{kw\_low} are particularly significant because they define the upper and lower limits of the search process, which in turn affects the size of the identified halos. In the actual calculation process, the computational resources used for large scales (small $k_w$) are minimal; thus, setting $k_w$ to 0.1 has little impact on the results, unless the simulation box is very large. Decreasing \texttt{kw\_high} reduces the number of small halos, but increasing \texttt{kw\_high} to a value that exceeds the simulation's resolution not only identifies unphysically small structures with particle numbers below \texttt{particle\_th} but also requires a significant amount of RAM. The impact of \texttt{n\_ref} on the HMF is similar to that of \texttt{kw\_high}, except that a significant decrease in the HMF is observed for larger particle numbers when \texttt{n\_ref} is less than 300.

With solely the HMF, it is not sufficient to deeply explore the effects of these parameters. To examine how these parameters affect the spatial distribution of halos, the HPS is calculated for these sets of catalogs. As shown in Figure~\ref{ps}, there is a noticeable shift between the HPS of the FOF and CWT catalogs. In the following discussion, we use the HPS of the FOF catalog, which has undergone the self-boundness check, for comparison. All four sets of HPS are depicted in Figure~\ref{PS_para}. Clearly, changes in these four parameters lead to a more significant difference in HPS compared to HMF.

In the case of \texttt{w\_resolution}, its effect is primarily limited to the HMF in the largest mass range, while its potential impact on the halo distribution is more clearly observed in the HPS. Reducing \texttt{w\_resolution} causes the HPS to shift towards lower values, with a more significant difference between these HPSs at smaller scales. This phenomenon can be attributed to the increased distance in the scale dimension between adjacent CWT slices, leading to a higher grid bias, as explained in Section~\ref{sec:gridbias}. Consequently, this leads to a change in the spatial distribution of halos. In contrast, increasing \texttt{w\_resolution} has minimal impact on the HPS, thus confirming that a \texttt{w\_resolution} of 20 is sufficient.

Similar to \texttt{w\_resolution}, the spatial resolution \texttt{n\_ref} has a significantly greater impact on the HPS. The underlying reasons for these phenomena are similar. Increasing \texttt{n\_ref} results in a greater difference between the grid resolutions of adjacent CWT slices, which in turn leads to a higher grid bias. Furthermore, improving the grid resolution consistently results in a reduction in halo size, as questionable boundary regions are excluded. Consequently, increasing \texttt{n\_ref} leads to a reduction in the total particle number within halos and a decline in the HPS.

The remaining two parameters, \texttt{kw\_high} and \texttt{dens\_th}, have only a negligible impact on the HPS. As previously mentioned, their effects are primarily limited to the smallest halos, which contribute minimally to the HPS calculation. It is important to note that this section does not provide default values for \texttt{kw\_high} and \texttt{kw\_low}. In actual calculations, these values should be determined based on the box length and resolution of the simulation. Moreover, the selection of these values should be guided by the size of the halos of interest to the user, as explained by Equation~(\ref{relation}).

\section{Conclusions and Discussions}
\label{sec:concl}

The present study introduces the CWTHF, a novel wavelet-based halo finder that is MPI parallelized and exhibits linear time complexity. The methodology of the CWTHF begins with the selection of appropriate identification parameters, where \texttt{kw\_low} and \texttt{kw\_high} determine the range of scales at which the CWTHF searches for halos. This approach has been shown to significantly impact the final catalog. By setting a low \texttt{kw\_high} value, the CWTHF can operate quickly and with low memory consumption, identifying only large halos. This enables the program to gain a preliminary understanding of particle clustering. The density threshold $\rho_{\rm th}=4$ restricts the grid density of halos, but its value is significantly smaller than that predicted by the spherical collapse model. This discrepancy can be attributed to the extension of the CWT boundary, which is used to define a more refined halo boundary. It is important to note that the remaining parameters serve only as resolution parameters, and a set of default values is pre-established for them.

After selecting the parameters, the CWT at each scale is calculated and segmented by the local maxima of the grid CWT. If a structure at this scale contains a maximum in the 4D wavelet space, particles are assigned to this structure and registered as a halo if they pass the self-boundness check. The identification capability of the CWTHF was tested by applying it to the SIMBA m50n512 dark matter sub-simulation and comparing the results with those of the FOF algorithm. The CWTHF identified fewer halos but contained more particles than the FOF, which contradicts the observations from the 2D test. The unbinding procedure applied to the FOF catalog reduced the discrepancy in the number of halos; however, the lack of a regathering process led to a sharp decline in the total particle number.

A test of the parameters within the CWTHF provides a set of default values for running the identification program. While the impact is considered negligible within the scope of the test, the upper limit of $k_w$, designated as \texttt{kw\_high}, determines the smallest structures that the CWTHF considers. Furthermore, it is observed that the usage of computational resources depends on the dynamic grid resolution in relation to $k_w$. Consequently, it is important to adjust \texttt{kw\_high} with caution when changes are made to the box length and particle number of the simulation. In addition to \texttt{kw\_high}, the spatial resolution \texttt{n\_ref} and the scale resolution \texttt{w\_resolution} have the most significant impact on the final catalog.

The properties of the CWT catalog and the effects of its parameters are summarized below:
\begin{enumerate}
  \item The spatial distribution of the CWT halos forms a web-like structure similar to that of the FOF, but it lacks the smallest halos in dense regions when compared to the distribution of the FOF halos.

  \item Unlike the previous 2D test, the CWT halos contain slightly more particles, but the number of halos is significantly lower than that of the FOF. This is attributed to the unbound nature of small FOF halos. By applying a self-boundness check, approximately 30\% of the FOF halos are excluded, reducing the excess in the FOF catalog to about 10\%.

  \item To achieve a smoother boundary, the CWTHF extends the boundary outward by removing the CWT value threshold and uses a self-boundness check to eliminate potentially unbound particles. This unbinding process successfully removes the jagged shape of the halo boundary, even when the spatial resolution is much lower than in the 2D test. Additionally, this transformation turns the CWTHF into a phase-space halo finder, as opposed to a position-space halo finder like the FOF.

  \item The differences between the CWT and FOF catalogs can be further examined using the HMF and HPS. The HMF indicates that the CWTHF identifies more medium to large halos and fewer small halos. The higher abundance of medium-sized halos is attributed to the splitting of large halos during the segmentation process. In the HPS, a shift to a higher amplitude is observed in the FOF, similar to the 2D test. The application of an unbinding process significantly reduces the bias caused by unphysically large halos in the FOF catalog.

  \item A test of the parameters provides a set of suggested parameter values. These values are set as the default in the CWTHF program. The selection of \texttt{kw\_high} should be carefully considered in relation to the box length and the simulation resolution.
\end{enumerate}

The identification results of the 3D CWTHF and the previous 2D test are generally consistent, except that the number of halos, particularly in the low particle number range, is lower in the 3D CWTHF than in the 2D test. Additionally, our research further confirms that the use of a self-boundness check significantly increases the consistency between different halo finders, as previously observed in \citet{Knebe2011, Knebe2013}.

Our program still has great potential for future optimizations. Even with the fast algorithm, the actual performance of the CWTHF has no obvious advantage over that of the FOF. In addition, the memory usage, which is associated with the \texttt{kw\_high}, depends on the size of the simulation box. For simulations with larger box lengths, it is necessary to set a higher \texttt{kw\_high} in order to access small-scale halos. This, of course, results in higher memory usage. Fortunately, an increase in box length generally corresponds to a decrease in spatial resolution. This enables us to avoid the necessity of extending the grid to a high value, thereby mitigating the problematic memory usage in these instances.

\section*{Acknowledgments}

The authors thank Dr. Miguel Aragon for his helpful comments and suggestions. MXL thanks anonymous users on Github for their assistance with some coding techniques. The authors also express their gratitude to the SIMBA team for making their data publicly available. The source code for CWTHF is available on Zenodo \citep{li_2025_15015510}\footnote{For future updated version of the CWTHF, see \url{https://github.com/salty-fish-114514/CWTHF/tree/main}.}

\software{NumPy \citep{vanderWalt2011,Harris2020}\footnote{\url{https://numpy.org/}}, SciPy \citep{Virtanen2020}\footnote{\url{https://scipy.org/}}, Matplotlib \citep{Hunter2007}\footnote{\url{https://matplotlib.org/}}, Jupyter Notebook\footnote{\url{https://jupyter.org/}}, Cython\footnote{\url{https://cython.org/}}, yt project \citep{Turk2011}\footnote{\url{https://github.com/yt-project}}}

\appendix

\section{Performance and Parallelization Efficiency}
\label{sec:performance}
\restartappendixnumbering

\begin{figure}[t]
    \centering
    \includegraphics[width=0.44\textwidth]{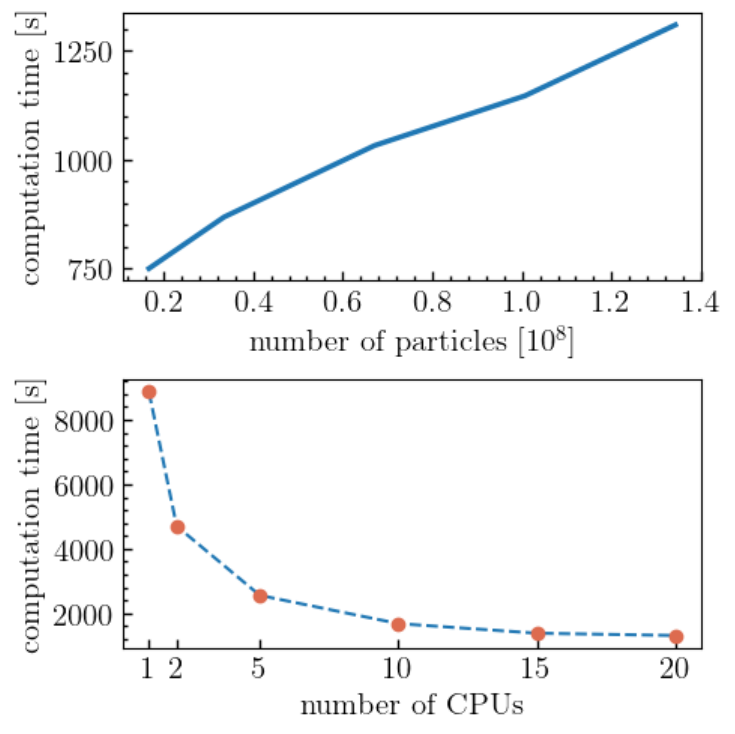}
    \caption{The computation time varies with different data sizes (top) and the number of CPUs (bottom). In all tests, the parameters are set to their default values.}
    \label{performance}
\end{figure}

The performance evaluation of the CWTHF includes two components: the actual computation time and the parallelization efficiency. Since the particle arrangement component of the CWTHF requires only a modest amount of computation time and the self-boundness check program is well established in other halo finders, this performance evaluation focuses solely on the parallelized part of the CWTHF, as detailed in Section~\ref{sec:Parallelization}. The performance tests are conducted on a platform equipped with dual Intel Xeon E5-2678 v3 CPUs. The results of these tests are shown in Figure~\ref{performance}. It is worth noting that the single-process FOF takes approximately 3 hours to construct a catalog, which is comparable to the time required by the single-process CWTHF.

The computational time required to process different data sizes demonstrates the effectiveness of the CWTHF in handling larger datasets. A dataset that is eight times larger results in only approximately 0.6 times more computation time. However, in actual calculations, increasing the particle number should be accompanied by an increase in \texttt{kw\_high} to identify smaller structures, which can lead to a decrease in performance. This evaluation further confirms that the time complexity of the CWTHF is of the order $\mathcal{O}(N)$.

The use of ghost cells has been shown to significantly reduce parallelization efficiency. Running the program on more than 10 CPUs does not lead to a noticeable increase in processing speed. Despite the reduced parallelization efficiency, the performance of the CWTHF is considered sufficient for practical applications, even in the more complex 3D space, thanks to the fast algorithm and the use of the Cython.

\section{Correction of Peak Grid Points}
\label{sec:correction}
\restartappendixnumbering

\begin{figure*}[t]
    \centering
    \includegraphics[width=0.98\textwidth]{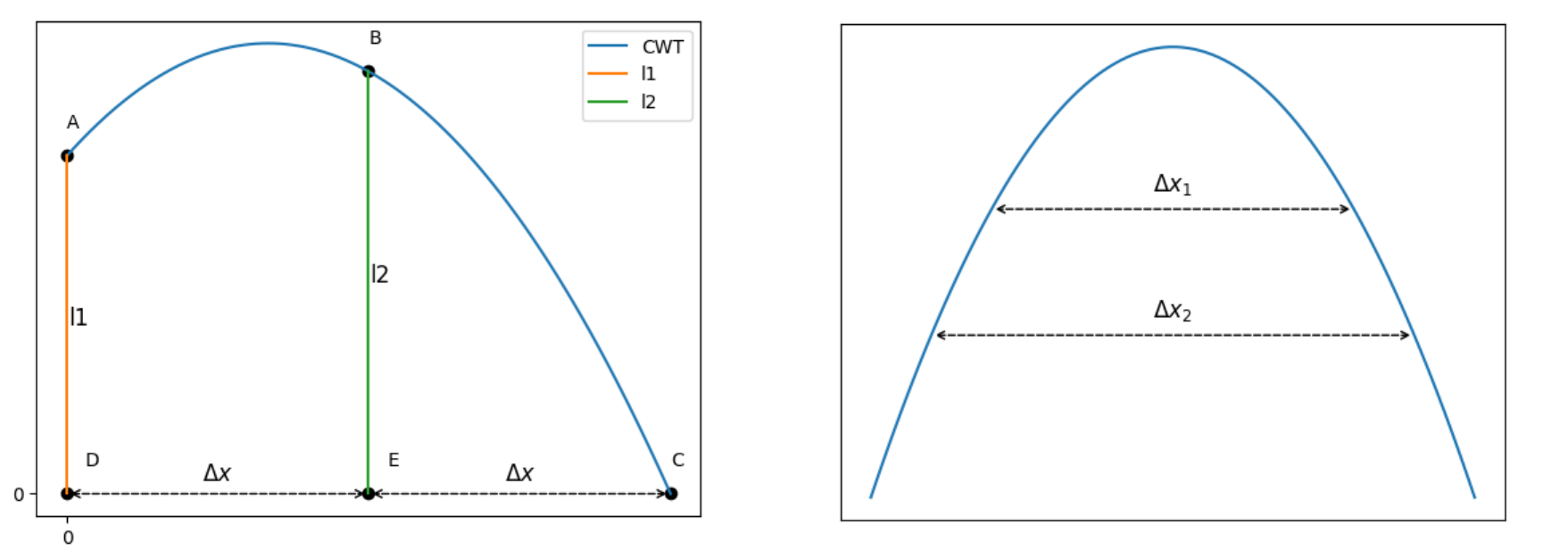}
    \caption{Left: The pattern of the discrepancy between the peak grid cube and the CWT peak. The grid maximum aligns with the value at grid point B, while the actual maximum of the CWT is higher than the value at B. Right: The pattern of bias resulting from different grid sizes. For a given peak, the mean value obtained from a smaller grid size $\Delta x_1$ is greater than that obtained from a larger grid size $\Delta x_2$.}
    \label{app_b}
\end{figure*}

Suppose the shape of the CWT near the peak can be approximated by a quadratic function. If the grid cube is located at the peak of the function, the values of the adjacent grid points are expected to be similar. However, in most cases, there is a discrepancy between the peak grid cube and the CWT peak, leading to the neighboring grid points having different values. The one-dimensional version of this pattern is shown in the left panel of Figure~\ref{app_b}.

Point B represents the maximum grid point that does not align with the CWT peak. Points A and C represent the neighboring grid points of this maximum in a specific dimension (i.e., on the x-axis). It should be noted that these three points can always be arranged as shown in the left panel of Figure~\ref{app_b}. The quadratic function passes through points A, B, and C, with their coordinates set to $(0,l_1)$, $(\Delta x,l_2)$, and $(2\Delta x,0)$, respectively.

In this coordinate system, the analytical form of the quadratic function can be derived relatively easily:
\begin{flalign}
    f(x) = \frac{l_1-2l_2}{2\Delta x ^2}(x-2\Delta x)\left(x-\frac{l_1\Delta x}{l_1-2l_2}\right).
\end{flalign}
The maximum value is given by the expression $(l_1-4l_2)^2/(16l_2-8l_1)$, and the correction factor for this dimension is $l_1^2/(16l_2-8l_1)$. Repeat this process for all three dimensions, adding the resulting correction factors to the maximum grid point. This ensures that the distance bias is minimized.

To further reduce the bias caused by different grid cube sizes when comparing maxima across scales, the maxima from the larger scale (with the larger grid size) are adjusted. The value at a given grid point represents the average value of the entire grid cube. The 1D example of this process is illustrated in the right panel of Figure~\ref{app_b}, where the value of the peak grid point is the average over $\Delta x_2$. When comparing this with the peak grid point of a smaller scale, the size of the smaller grid cube is $\Delta x_1$, which inherently corresponds to a higher average value even if the CWT is the same.

In the event that the CWT at the peak can be approximated by a 3D isotropic quadratic function, the grid peak value, designated as $V_{\rm pk}$, can be expressed as follows:
\begin{flalign}
 \label{pk}
 V_{\rm pk} =\frac{\int_0^{r_2/2} 4 \pi r^2(c-ar^2)dr}{4/3\pi (r_2/2)^3},
\end{flalign}
where $r_2/2=\Delta x_2/2$ represents the `radius' of the grid cube, $c$ and $a$ are parameters of the quadratic function. For simplicity, Equation~(\ref{pk}) is integrated over a spherical volume instead of a cubic volume
\begin{flalign}
  V_{\rm pk} =c+\frac{3}{5}a(r_2/2)^2.
\end{flalign}
Suppose the CWT value at $r=r_2/2$ to be $l$, then
\begin{flalign}
    c-a(r_2/2)^2=l.
\end{flalign}
By combining the equations mentioned earlier, the value of $a$ can be determined.
\begin{flalign}
  a=\frac{5}{2(r_2/2)^2}(V_{\rm pk}-l).
\end{flalign}
Assuming that the shape of this 3D function is simply the average of the three aforementioned 1D quadratic functions, the distance between $V_{\rm pk}$ and $l$ can be calculated as follows:
\begin{flalign}
  V_{\rm pk}-l&=-\frac{\overline{l}_1 - 2\overline{l}_2}{2\Delta x_2 ^2}(r_2/2)^2 \nonumber \\
  & = \frac{2\overline{l}_2-\overline{l}_1}{2\Delta x_2 ^2}(\Delta x_2/2)^2=\frac{2\overline{l}_2 - \overline{l}_1}{8},
\end{flalign}
where $\overline{l}_1$, $\overline{l}_2$ represent the averages of $l_1$, $l_2$ over all three dimensions.

The radius is now reduced to $r_1/2$, and the discrepancy between these two values is given by
\begin{flalign}
  &\frac{\int_0^{r_2/2} 4 \pi r^2(c-ar^2)dr}{4/3\pi (r_2/2)^3}-\frac{\int_0^{r_1/2} 4 \pi r^2(c-ar^2)dr}{4/3\pi (r_1/2)^3} \nonumber \\
  = & \frac{3}{5}a\left(\frac{r_2}{2}\right)^2 - \frac{3}{5}a\left(\frac{r_1}{2}\right)^2 \nonumber \\
  = & \frac{3}{5}\frac{5}{2(r_2/2)^2}\frac{2\overline{l}_2-\overline{l}_1}{8}\left[\left(\frac{r_2}{2}\right)^2-\left(\frac{r_1}{2}\right)^2\right]  \nonumber \\
  = & \frac{3}{16}(2\overline{l}_2 - \overline{l}_1)\left[1 - \left(\frac{r_1}{r_2}\right)^2\right]  \nonumber.
\end{flalign}
Nevertheless, the precise form of the CWT peak is sharper than that of a quadratic function. Consequently, the equation mentioned earlier underestimates the discrepancy between different resolutions. Removing the factor 3/5 will improve consistency across different resolutions in our program.

\bibliography{Ref}{}
\bibliographystyle{aasjournal}
\end{CJK*}
\end{document}

%% file: flowchart.tex
	\begin{figure}[t]  
		\scriptsize  
		\tikzstyle{format}=[rectangle,draw,thin,fill=blue!50!black!20, draw=none]  
		\tikzstyle{test}=[diamond,aspect=2,draw,thin,fill=blue!50!black!20, draw=none]  
		\tikzstyle{point}=[coordinate,on grid,]  
		\begin{tikzpicture}[scale=0.44]
		\node[rectangle,draw,thin,fill=blue!50!black!20,align=center, draw=none] (start){Initialize the program:\\
                                set the range of the $k_w$ \\
                                and scale resolution $N_w$};
		\node[format,below of=start,node distance=15mm](initi){$i = 1$};
  	\node[format,below of=initi,align=center,node distance=15mm](initgrid){
                                calculating\_grid = []\\
                                current\_grid = []};
		\node[test,below of=initgrid,node distance=20mm](circulate){$i \leq N_w$?};
		\node[format,left of=circulate,node distance=20mm](iplus1){
        $i=i+1$
        };
        \node[format,right of=circulate,node distance=40mm,align=center](last_iterate){
                                calculating\_grid = current\_grid\\
                                current\_grid = zero like array};
		\node[format,above of=last_iterate,align=center](haloseg2){Find all the local maxima in\\
                                calculating\_grid and segment the grid};     
		\node[format,above of=haloseg2,align=center,node distance=12mm](label2){Assign particles with no label\\
                                to the segmented grid and \\
                                label them temp ID};  \\ 
		\node[format,above of=label2,align=center](check2){
        Discard unphysical groups};    
		\node[format,above of=check2,align=center](updatelabel2){Store the particles that \\
                                 pass self-boundness check and\\
                                 label them with halo ID};   
		\node[rectangle,draw,thin,fill=blue!50!black!20,above of=updatelabel2,align=center, draw=none](end){Save labeled particles \\
                                 and stop the program};                                  
		\node[test,below of=circulate,node distance=15mm](checki){$i = 1$?};
		\node[format,below of=checki,node distance=15mm,align=center](setkw1){Select the $i$-th element \\
                                of $k_w$, $k_{w,i}$};
		\node[format,below of=setkw1,align=center](NgNw){$N_g$ = Int(400$k_{w,i}$+150)\\
                                $N_w$ = Int(1.5/$k_{w,i}$+5)};
		\node[format,below of=NgNw,align=center](gridCWT1){
        Assign particles to $N_g$\\
                                grid and calculate the grid CWT};
        \node[format,below of=gridCWT1,align=center](updategrid1){
                                calculating\_grid = current\_grid\\
                                current\_grid = ${\rm C}_i$};
		\node[point,below of=updategrid1,node distance=10mm](point1){};
        \node[point,left of=point1,node distance=20mm](point2){};
        \node[format,right of=checki,node distance=40mm,align=center](setkw2){Select the $i$-th element \\
                                of $k_w$, $k_{w,i}$};
		\node[format,below of=setkw2,align=center](NgNw2){$N_g$ = Int(400$k_{w,i}$+150)\\
                                $N_w$ = Int(1.5/$k_{w,i}$+5)};
		\node[format,below of=NgNw2,align=center](gridCWT2){
        Assign particles to $N_g$\\
                                grid and calculate the grid CWT};
        \node[format,below of=gridCWT2,align=center](updategrid2){
                                calculating\_grid = current\_grid\\
                                current\_grid = ${\rm C}_i$};
		\node[format,below of=updategrid2,align=center](haloseg){Find all the local maxima in\\
                                calculating\_grid and segment the grid};     
		\node[format,below of=haloseg,align=center,node distance=12mm](label){Assign particles with no label\\
                                to the segmented grid and \\
                                label them temp ID};  \\ 
		\node[format,below of=label,align=center](check){
        Discard unphysical groups};    
		\node[format,below of=check,align=center](updatelabel){Store the particles that \\
                                 pass self-boundness check and\\
                                 label them with halo ID};    
        \node[point,left of=updatelabel,node distance=60mm](point5){};
		\draw[->] (start)--(initi);
        \draw[->] (initi)--(initgrid);
        \draw[->] (initgrid)--(circulate);
		\draw[->](circulate) --node[left, text=green!60!black!90]{Yes} (checki);
        \draw[->](circulate) --node[above, text=red]{No} (last_iterate);
        \draw[->](last_iterate) -- (haloseg2);
        \draw[->](haloseg2)--(label2);
        \draw[->](label2)--(check2);
        \draw[->](check2)--(updatelabel2);
        \draw[->](updatelabel2)--(end);
		\draw[->](checki)--node[left, text=green!60!black!90]{Yes} (setkw1);
		\draw[->](setkw1)--(NgNw);
		\draw[->](NgNw)--(gridCWT1);
        \draw[->](gridCWT1)--(updategrid1);
        \draw[->, rounded corners=2mm] (updategrid1) -- (point1) -- (point2);
        \draw[->](point2)--(iplus1);
        \draw[->](iplus1)--(circulate);
        \draw[->](checki)--node[above, text=red]{No} (setkw2);
		\draw[->](setkw2)--(NgNw2);
		\draw[->](NgNw2)--(gridCWT2);
        \draw[->](gridCWT2)--(updategrid2);
        \draw[->](updategrid2)--(haloseg);
        \draw[->](haloseg)--(label);
        \draw[->](label)--(check);
        \draw[->](check)--(updatelabel);
        \draw[-, rounded corners=2mm](updatelabel)--(point5)--(point2);
		\end{tikzpicture}  
        \caption{The flowchart of the logical methodology of our CWTHF program. In this flowchart, we use the default values of the parameters as detailed in Section~\ref{sec:parameters}.}
        \label{approach_flowchart}
	\end{figure} 

%% file: main.bbl
\begin{thebibliography}{}
\expandafter\ifx\csname natexlab\endcsname\relax\def\natexlab#1{#1}\fi
\providecommand{\url}[1]{\href{#1}{#1}}
\providecommand{\dodoi}[1]{doi:~\href{http://doi.org/#1}{\nolinkurl{#1}}}
\providecommand{\doeprint}[1]{\href{http://ascl.net/#1}{\nolinkurl{http://ascl.net/#1}}}
\providecommand{\doarXiv}[1]{\href{https://arxiv.org/abs/#1}{\nolinkurl{https://arxiv.org/abs/#1}}}

\bibitem[{{Arag{\'o}n-Calvo} {et~al.}(2007){Arag{\'o}n-Calvo}, {Jones}, {van de Weygaert}, \& {van der Hulst}}]{Aragon2007}
{Arag{\'o}n-Calvo}, M.~A., {Jones}, B.~J.~T., {van de Weygaert}, R., \& {van der Hulst}, J.~M. 2007, \aap, 474, 315, \dodoi{10.1051/0004-6361:20077880}

\bibitem[{{Balaudo} {et~al.}(2024){Balaudo}, {Calore}, {De Romeri}, \& {Donato}}]{Balaudo2024}
{Balaudo}, A., {Calore}, F., {De Romeri}, V., \& {Donato}, F. 2024, \jcap, 2024, 001, \dodoi{10.1088/1475-7516/2024/02/001}

\bibitem[{{Baluev} \& {Rodionov}(2020)}]{Baluev2020}
{Baluev}, R.~V., \& {Rodionov}, E.~I. 2020, Celestial Mechanics and Dynamical Astronomy, 132, 34, \dodoi{10.1007/s10569-020-09976-2}

\bibitem[{{Behroozi} {et~al.}(2013){Behroozi}, {Wechsler}, \& {Wu}}]{Behroozi2013}
{Behroozi}, P.~S., {Wechsler}, R.~H., \& {Wu}, H.-Y. 2013, \apj, 762, 109, \dodoi{10.1088/0004-637X/762/2/109}

\bibitem[{{Bendjoya} {et~al.}(1991){Bendjoya}, {Slezak}, \& {Froeschle}}]{Bendjoya1991}
{Bendjoya}, P., {Slezak}, E., \& {Froeschle}, C. 1991, \aap, 251, 312

\bibitem[{{Bertschinger} \& {Gelb}(1991)}]{Bertschinger1991}
{Bertschinger}, E., \& {Gelb}, J.~M. 1991, Computers in Physics, 5, 164, \dodoi{10.1063/1.4822978}

\bibitem[{{Bijaoui} {et~al.}(1992){Bijaoui}, {Slezak}, \& {Mars}}]{Bijaoui1992}
{Bijaoui}, A., {Slezak}, E., \& {Mars}, G. 1992, in Distribution of Matter in the Universe, ed. G.~A. {Mamon} \& D.~{Gerbal}, 323--332

\bibitem[{{Cay{\'o}n} {et~al.}(2000){Cay{\'o}n}, {Sanz}, {Barreiro}, {Mart{\'\i}nez-Gonz{\'a}lez}, {Vielva}, {Toffolatti}, {Silk}, {Diego}, \& {Arg{\"u}eso}}]{Cayon2000}
{Cay{\'o}n}, L., {Sanz}, J.~L., {Barreiro}, R.~B., {et~al.} 2000, \mnras, 315, 757, \dodoi{10.1046/j.1365-8711.2000.03462.x}

\bibitem[{{Ciprini} {et~al.}(2007){Ciprini}, {Tosti}, {Marcucci}, {Cecchi}, {Discepoli}, {Bonamente}, {Germani}, {Impiombato}, {Lubrano}, \& {Pepe}}]{Ciprini2007}
{Ciprini}, S., {Tosti}, G., {Marcucci}, F., {et~al.} 2007, in American Institute of Physics Conference Series, Vol. 921, The First GLAST Symposium, ed. S.~{Ritz}, P.~{Michelson}, \& C.~A. {Meegan}, 546--547

\bibitem[{{Dav{\'e}} {et~al.}(2019){Dav{\'e}}, {Angl{\'e}s-Alc{\'a}zar}, {Narayanan}, {Li}, {Rafieferantsoa}, \& {Appleby}}]{Dave2019}
{Dav{\'e}}, R., {Angl{\'e}s-Alc{\'a}zar}, D., {Narayanan}, D., {et~al.} 2019, \mnras, 486, 2827, \dodoi{10.1093/mnras/stz937}

\bibitem[{{Davis} {et~al.}(1985){Davis}, {Efstathiou}, {Frenk}, \& {White}}]{Davis1985}
{Davis}, M., {Efstathiou}, G., {Frenk}, C.~S., \& {White}, S.~D.~M. 1985, \apj, 292, 371, \dodoi{10.1086/163168}

\bibitem[{{Djafer} {et~al.}(2012){Djafer}, {Irbah}, \& {Meftah}}]{Djafer2012}
{Djafer}, D., {Irbah}, A., \& {Meftah}, M. 2012, \solphys, 281, 863, \dodoi{10.1007/s11207-012-0109-3}

\bibitem[{{Eisenstein} \& {Hut}(1998)}]{Eisenstein1998}
{Eisenstein}, D.~J., \& {Hut}, P. 1998, \apj, 498, 137, \dodoi{10.1086/305535}

\bibitem[{{Flin} \& {Krywult}(2006)}]{Flin2006}
{Flin}, P., \& {Krywult}, J. 2006, \aap, 450, 9, \dodoi{10.1051/0004-6361:20041635}

\bibitem[{{Freeman} {et~al.}(2002){Freeman}, {Kashyap}, {Rosner}, \& {Lamb}}]{Freeman2002}
{Freeman}, P.~E., {Kashyap}, V., {Rosner}, R., \& {Lamb}, D.~Q. 2002, \apjs, 138, 185, \dodoi{10.1086/324017}

\bibitem[{{Frenk} \& {White}(2012)}]{Frenk2012}
{Frenk}, C.~S., \& {White}, S.~D.~M. 2012, Annalen der Physik, 524, 507, \dodoi{10.1002/andp.201200212}

\bibitem[{{Gelb} \& {Bertschinger}(1994)}]{Gelb1994}
{Gelb}, J.~M., \& {Bertschinger}, E. 1994, \apj, 436, 467, \dodoi{10.1086/174922}

\bibitem[{{Goetz} {et~al.}(1998){Goetz}, {Huchra}, \& {Brandenberger}}]{Goetz1998}
{Goetz}, M., {Huchra}, J.~P., \& {Brandenberger}, R.~H. 1998, arXiv e-prints, astro, \dodoi{10.48550/arXiv.astro-ph/9811393}

\bibitem[{{Gonz{\'a}lez-G{\'o}mez} {et~al.}(2010){Gonz{\'a}lez-G{\'o}mez}, {Blanco-Cano}, \& {Raga}}]{Gonzalez2010}
{Gonz{\'a}lez-G{\'o}mez}, D.~I., {Blanco-Cano}, X., \& {Raga}, A.~C. 2010, Advances in Space Research, 46, 22, \dodoi{10.1016/j.asr.2010.02.022}

\bibitem[{{Grebenev} {et~al.}(1995){Grebenev}, {Forman}, {Jones}, \& {Murray}}]{Grebenev1995}
{Grebenev}, S.~A., {Forman}, W., {Jones}, C., \& {Murray}, S. 1995, \apj, 445, 607, \dodoi{10.1086/175725}

\bibitem[{{Hadzhiyska} {et~al.}(2022){Hadzhiyska}, {Eisenstein}, {Bose}, {Garrison}, \& {Maksimova}}]{Hadzhiyska2022}
{Hadzhiyska}, B., {Eisenstein}, D., {Bose}, S., {Garrison}, L.~H., \& {Maksimova}, N. 2022, \mnras, 509, 501, \dodoi{10.1093/mnras/stab2980}

\bibitem[{{Harris} {et~al.}(2020){Harris}, {Millman}, {van der Walt}, {Gommers}, {Virtanen}, {Cournapeau}, {Wieser}, {Taylor}, {Berg}, {Smith}, {Kern}, {Picus}, {Hoyer}, {van Kerkwijk}, {Brett}, {Haldane}, {del R{\'\i}o}, {Wiebe}, {Peterson}, {G{\'e}rard-Marchant}, {Sheppard}, {Reddy}, {Weckesser}, {Abbasi}, {Gohlke}, \& {Oliphant}}]{Harris2020}
{Harris}, C.~R., {Millman}, K.~J., {van der Walt}, S.~J., {et~al.} 2020, \nat, 585, 357, \dodoi{10.1038/s41586-020-2649-2}

\bibitem[{{Hayn} {et~al.}(2012){Hayn}, {Panet}, {Diament}, {Holschneider}, {Mandea}, \& {Davaille}}]{Hayn2012}
{Hayn}, M., {Panet}, I., {Diament}, M., {et~al.} 2012, Geophysical Journal International, 189, 1430, \dodoi{10.1111/j.1365-246X.2012.05455.x}

\bibitem[{{Hopkins}(2015)}]{Hopkins2015}
{Hopkins}, P.~F. 2015, \mnras, 450, 53, \dodoi{10.1093/mnras/stv195}

\bibitem[{{Hunter}(2007)}]{Hunter2007}
{Hunter}, J.~D. 2007, Computing in Science and Engineering, 9, 90, \dodoi{10.1109/MCSE.2007.55}

\bibitem[{{Knebe} {et~al.}(2011){Knebe}, {Knollmann}, {Muldrew}, {Pearce}, {Aragon-Calvo}, {Ascasibar}, {Behroozi}, {Ceverino}, {Colombi}, {Diemand}, {Dolag}, {Falck}, {Fasel}, {Gardner}, {Gottl{\"o}ber}, {Hsu}, {Iannuzzi}, {Klypin}, {Luki{\'c}}, {Maciejewski}, {McBride}, {Neyrinck}, {Planelles}, {Potter}, {Quilis}, {Rasera}, {Read}, {Ricker}, {Roy}, {Springel}, {Stadel}, {Stinson}, {Sutter}, {Turchaninov}, {Tweed}, {Yepes}, \& {Zemp}}]{Knebe2011}
{Knebe}, A., {Knollmann}, S.~R., {Muldrew}, S.~I., {et~al.} 2011, \mnras, 415, 2293, \dodoi{10.1111/j.1365-2966.2011.18858.x}

\bibitem[{{Knebe} {et~al.}(2013){Knebe}, {Pearce}, {Lux}, {Ascasibar}, {Behroozi}, {Casado}, {Moran}, {Diemand}, {Dolag}, {Dominguez-Tenreiro}, {Elahi}, {Falck}, {Gottl{\"o}ber}, {Han}, {Klypin}, {Luki{\'c}}, {Maciejewski}, {McBride}, {Merch{\'a}n}, {Muldrew}, {Neyrinck}, {Onions}, {Planelles}, {Potter}, {Quilis}, {Rasera}, {Ricker}, {Roy}, {Ruiz}, {Sgr{\'o}}, {Springel}, {Stadel}, {Sutter}, {Tweed}, \& {Zemp}}]{Knebe2013}
{Knebe}, A., {Pearce}, F.~R., {Lux}, H., {et~al.} 2013, \mnras, 435, 1618, \dodoi{10.1093/mnras/stt1403}

\bibitem[{{Knollmann} \& {Knebe}(2009)}]{Knollmann2009}
{Knollmann}, S.~R., \& {Knebe}, A. 2009, \apjs, 182, 608, \dodoi{10.1088/0067-0049/182/2/608}

\bibitem[{{Kuhlen} {et~al.}(2012){Kuhlen}, {Vogelsberger}, \& {Angulo}}]{Kuhlen2012}
{Kuhlen}, M., {Vogelsberger}, M., \& {Angulo}, R. 2012, Physics of the Dark Universe, 1, 50, \dodoi{10.1016/j.dark.2012.10.002}

\bibitem[{{Lacey} \& {Cole}(1994)}]{Lacey1994}
{Lacey}, C., \& {Cole}, S. 1994, \mnras, 271, 676, \dodoi{10.1093/mnras/271.3.676}

\bibitem[{{Lazzati} {et~al.}(1999){Lazzati}, {Campana}, {Rosati}, {Panzera}, \& {Tagliaferri}}]{Lazzati1999}
{Lazzati}, D., {Campana}, S., {Rosati}, P., {Panzera}, M.~R., \& {Tagliaferri}, G. 1999, \apj, 524, 414, \dodoi{10.1086/307788}

\bibitem[{Li(2025)}]{li_2025_15015510}
Li, M. 2025, CWTHF (Continuous Wavelet Transform Halo Finder),  Zenodo, \dodoi{10.5281/zenodo.15015510}.
\newblock \url{https://doi.org/10.5281/zenodo.15015510}

\bibitem[{{Li} {et~al.}(2024){Li}, {Wang}, \& {He}}]{Li2024}
{Li}, M., {Wang}, Y., \& {He}, P. 2024, \apj, 973, 39, \dodoi{10.3847/1538-4357/ad65d1}

\bibitem[{{Li} \& {He}(2018)}]{Li2018}
{Li}, Z.-W., \& {He}, P. 2018, Communications in Theoretical Physics, 70, 728, \dodoi{10.1088/0253-6102/70/6/728}

\bibitem[{{Mertens} \& {Lobanov}(2015)}]{Mertens2015}
{Mertens}, F., \& {Lobanov}, A. 2015, \aap, 574, A67, \dodoi{10.1051/0004-6361/201424566}

\bibitem[{{Nelson} {et~al.}(2019){Nelson}, {Springel}, {Pillepich}, {Rodriguez-Gomez}, {Torrey}, {Genel}, {Vogelsberger}, {Pakmor}, {Marinacci}, {Weinberger}, {Kelley}, {Lovell}, {Diemer}, \& {Hernquist}}]{Nelson2019}
{Nelson}, D., {Springel}, V., {Pillepich}, A., {et~al.} 2019, Computational Astrophysics and Cosmology, 6, 2, \dodoi{10.1186/s40668-019-0028-x}

\bibitem[{{Patrikeev} {et~al.}(2006){Patrikeev}, {Fletcher}, {Stepanov}, {Beck}, {Berkhuijsen}, {Frick}, \& {Horellou}}]{Patrikeev2006}
{Patrikeev}, I., {Fletcher}, A., {Stepanov}, R., {et~al.} 2006, \aap, 458, 441, \dodoi{10.1051/0004-6361:20065225}

\bibitem[{{Peebles}(1974)}]{Peebles1974}
{Peebles}, P.~J.~E. 1974, \apjl, 189, L51, \dodoi{10.1086/181462}

\bibitem[{{Planck Collaboration} {et~al.}(2016){Planck Collaboration}, {Ade}, {Aghanim}, {Arnaud}, {Ashdown}, {Aumont}, {Baccigalupi}, {Banday}, {Barreiro}, {Bartlett}, {Bartolo}, {Battaner}, {Battye}, {Benabed}, {Beno{\^\i}t}, {Benoit-L{\'e}vy}, {Bernard}, {Bersanelli}, {Bielewicz}, {Bock}, {Bonaldi}, {Bonavera}, {Bond}, {Borrill}, {Bouchet}, {Boulanger}, {Bucher}, {Burigana}, {Butler}, {Calabrese}, {Cardoso}, {Catalano}, {Challinor}, {Chamballu}, {Chary}, {Chiang}, {Chluba}, {Christensen}, {Church}, {Clements}, {Colombi}, {Colombo}, {Combet}, {Coulais}, {Crill}, {Curto}, {Cuttaia}, {Danese}, {Davies}, {Davis}, {de Bernardis}, {de Rosa}, {de Zotti}, {Delabrouille}, {D{\'e}sert}, {Di Valentino}, {Dickinson}, {Diego}, {Dolag}, {Dole}, {Donzelli}, {Dor{\'e}}, {Douspis}, {Ducout}, {Dunkley}, {Dupac}, {Efstathiou}, {Elsner}, {En{\ss}lin}, {Eriksen}, {Farhang}, {Fergusson}, {Finelli}, {Forni}, {Frailis}, {Fraisse}, {Franceschi}, {Frejsel}, {Galeotta}, {Galli}, {Ganga}, {Gauthier}, {Gerbino}, {Ghosh}, {Giard},
  {Giraud-H{\'e}raud}, {Giusarma}, {Gjerl{\o}w}, {Gonz{\'a}lez-Nuevo}, {G{\'o}rski}, {Gratton}, {Gregorio}, {Gruppuso}, {Gudmundsson}, {Hamann}, {Hansen}, {Hanson}, {Harrison}, {Helou}, {Henrot-Versill{\'e}}, {Hern{\'a}ndez-Monteagudo}, {Herranz}, {Hildebrandt}, {Hivon}, {Hobson}, {Holmes}, {Hornstrup}, {Hovest}, {Huang}, {Huffenberger}, {Hurier}, {Jaffe}, {Jaffe}, {Jones}, {Juvela}, {Keih{\"a}nen}, {Keskitalo}, {Kisner}, {Kneissl}, {Knoche}, {Knox}, {Kunz}, {Kurki-Suonio}, {Lagache}, {L{\"a}hteenm{\"a}ki}, {Lamarre}, {Lasenby}, {Lattanzi}, {Lawrence}, {Leahy}, {Leonardi}, {Lesgourgues}, {Levrier}, {Lewis}, {Liguori}, {Lilje}, {Linden-V{\o}rnle}, {L{\'o}pez-Caniego}, {Lubin}, {Mac{\'\i}as-P{\'e}rez}, {Maggio}, {Maino}, {Mandolesi}, {Mangilli}, {Marchini}, {Maris}, {Martin}, {Martinelli}, {Mart{\'\i}nez-Gonz{\'a}lez}, {Masi}, {Matarrese}, {McGehee}, {Meinhold}, {Melchiorri}, {Melin}, {Mendes}, {Mennella}, {Migliaccio}, {Millea}, {Mitra}, {Miville-Desch{\^e}nes}, {Moneti}, {Montier}, {Morgante}, {Mortlock},
  {Moss}, {Munshi}, {Murphy}, {Naselsky}, {Nati}, {Natoli}, {Netterfield}, {N{\o}rgaard-Nielsen}, {Noviello}, {Novikov}, {Novikov}, {Oxborrow}, {Paci}, {Pagano}, {Pajot}, {Paladini}, {Paoletti}, {Partridge}, {Pasian}, {Patanchon}, {Pearson}, {Perdereau}, {Perotto}, {Perrotta}, {Pettorino}, {Piacentini}, {Piat}, {Pierpaoli}, {Pietrobon}, {Plaszczynski}, {Pointecouteau}, {Polenta}, {Popa}, {Pratt}, {Pr{\'e}zeau}, {Prunet}, {Puget}, {Rachen}, {Reach}, {Rebolo}, {Reinecke}, {Remazeilles}, {Renault}, {Renzi}, {Ristorcelli}, {Rocha}, {Rosset}, {Rossetti}, {Roudier}, {Rouill{\'e} d'Orfeuil}, {Rowan-Robinson}, {Rubi{\~n}o-Mart{\'\i}n}, {Rusholme}, {Said}, {Salvatelli}, {Salvati}, {Sandri}, {Santos}, {Savelainen}, {Savini}, {Scott}, {Seiffert}, {Serra}, {Shellard}, {Spencer}, {Spinelli}, {Stolyarov}, {Stompor}, {Sudiwala}, {Sunyaev}, {Sutton}, {Suur-Uski}, {Sygnet}, {Tauber}, {Terenzi}, {Toffolatti}, {Tomasi}, {Tristram}, {Trombetti}, {Tucci}, {Tuovinen}, {T{\"u}rler}, {Umana}, {Valenziano}, {Valiviita}, {Van Tent},
  {Vielva}, {Villa}, {Wade}, {Wandelt}, {Wehus}, {White}, {White}, {Wilkinson}, {Yvon}, {Zacchei}, \& {Zonca}}]{Planck2016}
{Planck Collaboration}, {Ade}, P.~A.~R., {Aghanim}, N., {et~al.} 2016, \aap, 594, A13, \dodoi{10.1051/0004-6361/201525830}

\bibitem[{{Planelles} \& {Quilis}(2010)}]{Planelles2010}
{Planelles}, S., \& {Quilis}, V. 2010, \aap, 519, A94, \dodoi{10.1051/0004-6361/201014214}

\bibitem[{{Press} \& {Schechter}(1974)}]{Press1974}
{Press}, W.~H., \& {Schechter}, P. 1974, \apj, 187, 425, \dodoi{10.1086/152650}

\bibitem[{{Romeo} {et~al.}(2008){Romeo}, {Agertz}, {Moore}, \& {Stadel}}]{Romeo2008}
{Romeo}, A.~B., {Agertz}, O., {Moore}, B., \& {Stadel}, J. 2008, \apj, 686, 1, \dodoi{10.1086/591236}

\bibitem[{{Slezak} {et~al.}(1993){Slezak}, {de Lapparent}, \& {Bijaoui}}]{Slezak1993}
{Slezak}, E., {de Lapparent}, V., \& {Bijaoui}, A. 1993, \apj, 409, 517, \dodoi{10.1086/172683}

\bibitem[{{Springel} {et~al.}(2001){Springel}, {White}, {Tormen}, \& {Kauffmann}}]{Springel2001}
{Springel}, V., {White}, S. D.~M., {Tormen}, G., \& {Kauffmann}, G. 2001, \mnras, 328, 726, \dodoi{10.1046/j.1365-8711.2001.04912.x}

\bibitem[{{Stadel}(2001)}]{Stadel2001}
{Stadel}, J.~G. 2001, PhD thesis, University of Washington, Seattle

\bibitem[{{Turk} {et~al.}(2011){Turk}, {Smith}, {Oishi}, {Skory}, {Skillman}, {Abel}, \& {Norman}}]{Turk2011}
{Turk}, M.~J., {Smith}, B.~D., {Oishi}, J.~S., {et~al.} 2011, \apjs, 192, 9, \dodoi{10.1088/0067-0049/192/1/9}

\bibitem[{{Vall{\'e}s-P{\'e}rez} {et~al.}(2022){Vall{\'e}s-P{\'e}rez}, {Planelles}, \& {Quilis}}]{Valles2022}
{Vall{\'e}s-P{\'e}rez}, D., {Planelles}, S., \& {Quilis}, V. 2022, \aap, 664, A42, \dodoi{10.1051/0004-6361/202243712}

\bibitem[{{van der Walt} {et~al.}(2011){van der Walt}, {Colbert}, \& {Varoquaux}}]{vanderWalt2011}
{van der Walt}, S., {Colbert}, S.~C., \& {Varoquaux}, G. 2011, Computing in Science and Engineering, 13, 22, \dodoi{10.1109/MCSE.2011.37}

\bibitem[{{Vavilova} \& {Babyk}(2018)}]{Vavilova2018}
{Vavilova}, I.~B., \& {Babyk}, I.~V. 2018, Odessa Astronomical Publications, 30, 239, \dodoi{10.18524/1810-4215.2018.31.146678}

\bibitem[{{Villaescusa-Navarro} {et~al.}(2020){Villaescusa-Navarro}, {Hahn}, {Massara}, {Banerjee}, {Delgado}, {Ramanah}, {Charnock}, {Giusarma}, {Li}, {Allys}, {Brochard}, {Uhlemann}, {Chiang}, {He}, {Pisani}, {Obuljen}, {Feng}, {Castorina}, {Contardo}, {Kreisch}, {Nicola}, {Alsing}, {Scoccimarro}, {Verde}, {Viel}, {Ho}, {Mallat}, {Wandelt}, \& {Spergel}}]{Villaescusa2020}
{Villaescusa-Navarro}, F., {Hahn}, C., {Massara}, E., {et~al.} 2020, \apjs, 250, 2, \dodoi{10.3847/1538-4365/ab9d82}

\bibitem[{{Virtanen} {et~al.}(2020){Virtanen}, {Gommers}, {Oliphant}, {Haberland}, {Reddy}, {Cournapeau}, {Burovski}, {Peterson}, {Weckesser}, {Bright}, {van der Walt}, {Brett}, {Wilson}, {Millman}, {Mayorov}, {Nelson}, {Jones}, {Kern}, {Larson}, {Carey}, {Polat}, {Feng}, {Moore}, {VanderPlas}, {Laxalde}, {Perktold}, {Cimrman}, {Henriksen}, {Quintero}, {Harris}, {Archibald}, {Ribeiro}, {Pedregosa}, {van Mulbregt}, \& {SciPy 1. 0 Contributors}}]{Virtanen2020}
{Virtanen}, P., {Gommers}, R., {Oliphant}, T.~E., {et~al.} 2020, Nature Methods, 17, 261, \dodoi{10.1038/s41592-019-0686-2}

\bibitem[{{Wang} {et~al.}(2008){Wang}, {Rowan-Robinson}, {Yamamura}, {Shibai}, {Savage}, {Oliver}, {Thomson}, {Rahman}, {Clements}, {Figueredo}, {Goto}, {Hasegawa}, {Jeong}, {Matsuura}, {M{\"u}ller}, {Nakagawa}, {Pearson}, {Serjeant}, {Shirahata}, \& {White}}]{Wang2008}
{Wang}, L., {Rowan-Robinson}, M., {Yamamura}, I., {et~al.} 2008, \mnras, 387, 601, \dodoi{10.1111/j.1365-2966.2008.13292.x}

\bibitem[{{Wang} \& {He}(2021)}]{Wang2021}
{Wang}, Y., \& {He}, P. 2021, Communications in Theoretical Physics, 73, 095402, \dodoi{10.1088/1572-9494/ac10be}

\bibitem[{{Wang} \& {He}(2022)}]{Wang2022b}
---. 2022, \apj, 934, 112, \dodoi{10.3847/1538-4357/ac7a3d}

\bibitem[{{Wang} \& {He}(2023)}]{Wang2023}
---. 2023, RAS Techniques and Instruments, 2, 307, \dodoi{10.1093/rasti/rzad020}

\bibitem[{{Wang} \& {He}(2024{\natexlab{a}})}]{Wang2024a}
---. 2024{\natexlab{a}}, \mnras, 528, 3797, \dodoi{10.1093/mnras/stae229}

\bibitem[{{Wang} \& {He}(2024{\natexlab{b}})}]{Wang2024b}
---. 2024{\natexlab{b}}, \apj, 974, 107, \dodoi{10.3847/1538-4357/ad6d63}

\bibitem[{{Wang} \& {He}(2024{\natexlab{c}})}]{Wang2024c}
---. 2024{\natexlab{c}}, \mnras, 534, L14, \dodoi{10.1093/mnrasl/slae073}

\bibitem[{{Wang} \& {He}(2024{\natexlab{d}})}]{Wang2025a}
---. 2024{\natexlab{d}}, arXiv e-prints, arXiv:2408.13876, \dodoi{10.48550/arXiv.2408.13876}

\bibitem[{{Wang} {et~al.}(2022){Wang}, {Yang}, \& {He}}]{Wang2022a}
{Wang}, Y., {Yang}, H.-Y., \& {He}, P. 2022, \apj, 934, 77, \dodoi{10.3847/1538-4357/ac752c}

\bibitem[{{Weinberg} {et~al.}(1997){Weinberg}, {Hernquist}, \& {Katz}}]{Weinberg1997}
{Weinberg}, D.~H., {Hernquist}, L., \& {Katz}, N. 1997, \apj, 477, 8, \dodoi{10.1086/303683}

\end{thebibliography}
